\documentclass[sigconf]{acmart}
\AtBeginDocument{%
  \providecommand\BibTeX{{%
    \normalfont B\kern-0.5em{\scshape i\kern-0.25em b}\kern-0.8em\TeX}}}

\usepackage{multirow}
\usepackage[skip=0.3\baselineskip]{caption}
\usepackage{enumitem}
\usepackage{stfloats}
\usepackage{wrapfig}

\copyrightyear{2023}
\acmYear{2023}
\setcopyright{acmlicensed}\acmConference[CHI '23]{Proceedings of the 2023 CHI Conference on Human Factors in Computing Systems}{April 23--28, 2023}{Hamburg, Germany}
\acmBooktitle{Proceedings of the 2023 CHI Conference on Human Factors in Computing Systems (CHI '23), April 23--28, 2023, Hamburg, Germany}
\acmPrice{15.00}
\acmDOI{10.1145/3544548.3581266}
\acmISBN{978-1-4503-9421-5/23/04}

\definecolor{cb_orange}{rgb}{1.0,0.51,0.0}
\definecolor{cb_blue}{rgb}{0, 0, 0}
\definecolor{cb_green}{rgb}{0.3,0.67,0.29}
\definecolor{cb_red}{rgb}{0.89,0.1,0.11}
\definecolor{cb_purple}{rgb}{0.6, 0.31, 0.64}
\definecolor{cb_brown}{rgb}{0.6, 0.4, 0.2}
\definecolor{cb_crimson}{rgb}{0.86, 0.08, 0.24}

\makeatletter
\def\subsubsection{\@startsection{subsubsection}{3}%
  \z@{.5\linespacing\@plus.7\linespacing}{.1\linespacing}%
  {\normalfont\itshape}}
\makeatother

\makeatletter
\renewcommand{\paragraph}{\@startsection{paragraph}{4}{\parindent}
  {-.1\baselineskip \@plus -2\p@ \@minus -.2\p@}%
  {-3.5\p@}%
  {\ACM@NRadjust{\@parfont\@adddotafter}}
}
\makeatother

\DeclareRobustCommand{\inlinefig}[1]{%
\begingroup
\setbox0=\hbox{\includegraphics[height=0.95em]{#1}}%
\parbox[c][10pt][t]{\wd0}{\box0}\endgroup
}

\newcommand{\cmo}[1]{{\textcolor{cb_blue}{#1}}}
\newcommand{\quot}[1]{\emph{``#1''}}

\newcommand{\COne}[0]{Casual fans are unsure about which players they should focus on.}
\newcommand{\CTwo}[0]{Casual fans are confused about the in-game decisions of players.}
\newcommand{\CThree}[0]{Casual fans have trouble seeking customized data while watching game videos.}

\newcommand{\DefS}[0]{Defense Shield}
\newcommand{\OffR}[0]{Offense Ring}
\newcommand{\OOOL}[0]{One-on-one Line}
\newcommand{\GFo}[0]{Gaze Focus}
\newcommand{\GFi}[0]{Gaze Filter}

\newcommand{\RAW}[0]{\textsf{RAW}}
\newcommand{\AUG}[0]{\textsf{AUG}}
\newcommand{\FULL}[0]{\textsf{FULL}}

\newcommand{\system}{iBall}

\begin{document}

\title{\system{}: Augmenting Basketball Videos with Gaze-moderated Embedded Visualizations}

\makeatletter
\let\@authorsaddresses\@empty
\makeatother

\author{Chen Zhu-Tian}
\orcid{0000-0002-2313-0612}
\affiliation{%
  \institution{Harvard University}
  \city{Cambridge, MA}
  \country{USA}
}
\email{ztchen@g.harvard.edu}

\author{Qisen Yang}
\authornote{
This work was done when Qisen Yang
and Jerry Shan
were interns at Harvard University.
They contributed equally to this research. }
\orcid{0000-0002-8378-735X}
\affiliation{%
  \institution{Zhejiang University}
  \city{Hangzhou, Zhejiang}
  \country{China}
}
\email{qs\_yang@zju.edu.cn}

\author{Jerry Shan}
\authornotemark[1]
\orcid{0000-0002-2304-2129}
\affiliation{%
  \institution{UC Berkeley}
  \city{Berkeley, CA}
  \country{USA}
}
\email{jiarui.shan@berkeley.edu}

\author{Tica Lin}
\orcid{0000-0002-2860-0871}
\affiliation{%
  \institution{Harvard University}
  \city{Cambridge, MA}
  \country{USA}
}
\email{tlin@g.harvard.edu}

\author{Johanna Beyer}
\orcid{0000-0002-3505-9171}
\affiliation{%
  \institution{Harvard University}
  \city{Cambridge, MA}
  \country{USA}
}
\email{jbeyer@g.harvard.edu}

\author{Haijun Xia}
\orcid{0000-0002-9425-0881}
\affiliation{%
  \institution{UC San Diego}
  \city{La Jolla, CA}
  \country{USA}
}
\email{haijunxia@ucsd.edu}

\author{Hanspeter Pfister}
\orcid{0000-0002-3620-2582}
\affiliation{%
  \institution{Harvard University}
  \city{Cambridge, MA}
  \country{USA}
}
\email{pfister@g.harvard.edu}

\begin{abstract}

We present \system{}, a basketball video-watching system that leverages gaze-moderated embedded visualizations to facilitate game understanding and engagement of casual fans. 
\cmo{Video broadcasting and online video platforms} make watching basketball games increasingly accessible. 
Yet, for new or casual fans, watching basketball videos is often confusing due to their limited basketball knowledge and the lack of accessible, on-demand information to resolve their confusion. 
To assist casual fans in watching basketball videos, we compared the game-watching behaviors of casual and die-hard fans in a formative study and developed \system{} based on the findings. 
\system{} embeds visualizations into basketball videos using a computer vision pipeline, 
and automatically adapts the visualizations based on the game context and users’ gaze,  helping casual fans appreciate basketball games without being overwhelmed. 
We confirmed the usefulness, usability, and engagement of \system{} in a study with 16 casual fans,
\cmo{and further collected feedback from 8 die-hard fans.}

\end{abstract}




\begin{teaserfigure}
  \hspace*{0.8cm}\includegraphics[width=\textwidth]{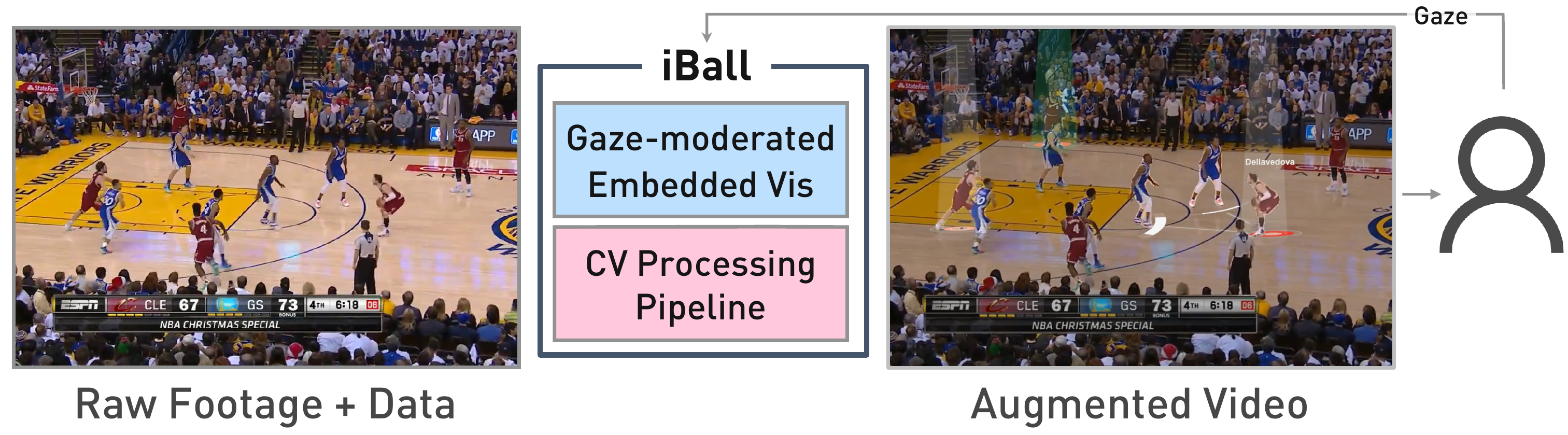}\hspace*{-2cm}
  \caption{\system{} augments basketball videos with 
    gaze-moderated embedded visualizations to facilitate game
    understanding and engagement of casual fans.
    It embeds data visualizations into basketball raw footage using a well-designed computer vision pipeline, 
  and automatically adapts the visualizations based on the game context and users' gaze.
  }
  \label{fig:teaser}
\end{teaserfigure}

\maketitle

\section{Introduction}
Basketball, one of the most popular team sports,
has continued to attract new fans
over the past decades due to the proliferation of \cmo{video broadcasting and online video platforms}.
However, as our formative study will show,
unlike experienced fans, 
new or casual fans often get confused
when watching basketball videos.
\cmo{
This is because they 
lack sufficient basketball knowledge 
to understand the players' complex teamwork and in-game decisions.
Existing methods of providing extra information, 
such as scoreboards in broadcasting videos and online webpages (e.g., ESPN~\cite{ESPN}), 
often fail to adequately address their confusion. 
These methods either cannot provide on-demand data or distract fans from the game by showing data in separate windows.}
Thus,
finding a way to enable seamless access to extra data 
while watching basketball games would particularly benefit casual fans 
in their understanding and engagement of games.

Embedded visualizations provide a promising opportunity
to allow audiences 
to access extra data 
without being distracted from the game video
by directly displaying data in the actual scenes.
Given such benefits,
various commercial products~\cite{vizrt, courtvision}
and research systems~\cite{VisCommentator, chen2022sporthesia}
leverage embedded visualizations 
to augment sports videos.
However, these systems either 
focus on post-game analysis rather than game-watching scenarios,
or only use simple, non-interactive text labels and progress bars to show data.
Recently, 
researchers have started to explore the design space of embedded visualizations for \cmo{game-watching scenarios}
but used low-fidelity simulated environments
(e.g., 3D simulated sports games~\cite{lin2022the}, moving charts on white backgrounds~\cite{yao2022visualization}).
Little is known about 
how to design and implement 
interactive embedded visualizations in real sports videos
and how they can facilitate game understanding and increase engagement of fans when watching games.

In this work,
we aim to fill this gap by developing interactive embedded visualizations
to assist casual fans in watching basketball game videos.
To understand the particular practices, pain points, and solutions of casual fans
in watching basketball game videos, 
we compared the game-watching behaviors of 8 casual and 8 die-hard fans
in a formative study.
Findings revealed that the casual fans were confused
about key players and their in-game decisions from time to time,
and \cmo{had trouble seeking customized data}
during the game.
Informed by the study, 
we developed \system{} (\autoref{fig:teaser}), 
a basketball game viewing system 
that 
automatically highlights key players
and visualizes their performance
through gaze-moderated embedded visualizations.

We developed \system{} by tackling two main challenges.
First, 
embedding visualizations into actual scenes 
is recognized as a grand challenge~\cite{DBLP:conf/chi/EnsBCESWPABDDGH21},
especially for basketball, where players overlap heavily
and the camera moves rapidly.
To tackle this challenge,
we contribute a CV pipeline that \cmo{pre-processes} team sports videos
for embedding visualizations.
\cmo{We also conducted experiments to evaluate our pipeline quantitatively and discuss potential methods to extend the pipeline to process live videos.}
Second, 
it remains unclear how to design embedded visualizations that
are informative but not overwhelming for individual audiences,
who may have various levels of game literacy, data needs, and personal interests.
We designed a set of gaze-moderated embedded visualizations
that leverage the user's gaze
to seamlessly 
present the data the user is interested in
and suppress others.
To evaluate \system{},
we conducted a user study with 16 casual fans
to compare the game-watching experiences 
between raw video (\RAW{}), 
video + embedded visualizations (\AUG{}), 
and video + gaze-moderated embedded visualizations (\FULL{}).
Participants spoke highly of our system,
ranked \FULL{} as the best, 
and confirmed that our embedded visualizations and gaze interactions
were useful, usable, and engaging. \cmo{We further collected and discuss the feedback on \system{} from another 8 die-hard fans.}
We discuss our observations and design implications learned from the study for future research inspiration.

In summary,
through developing \system{},
we make the following four main contributions:
1) a formative study that identifies the pain points of casual fans in watching basketball videos
and solicits plausible solutions from die-hard fans,
2) an open-source CV pipeline to process team sports videos for embedding visualizations, 
3) a set of gaze-moderated embedded visualizations for basketball game videos, and
4) a user study that assesses our system and provides 
insightful feedback on using gaze-moderated embedded visualizations in team sports videos.
Finally, we will open source our system at {\color{blue}\url{https://github.com/ASportsV/iBall}}. 
\section{Related Work}
We review prior work on personalized game viewing systems,
embedded visualizations in sports videos,
computer vision for embedded visualizations,
and applications of gaze interactions.

\subsection{Personalized Game Viewing Systems}
Visualization has long been used in sports to present data~\cite{PerinVSSWC18},
including box scores~\cite{DBLP:conf/chi/FuS22}, 
tracking data~\cite{SoccerStories, BKViz, ShuttleSpace, TIVEE}, 
and metadata~\cite{DBLP:journals/cga/VuillemotP16}.
Sports visualizations are mainly used for post-game analysis or in-game informing purposes.
This work mainly focuses on the latter.

Sports games usually involve complex in-game decision-making.
To better understand, analyze, and appreciate players' in-game decisions,
spectators often look for additional information when watching a sports game~\cite{lin2022the}.
To fulfill individual spectators' information needs, prior research has 
explored the design of interactive game-watching systems.
ARSpectator~\cite{DBLP:conf/siggrapha/ZollmannLLLB19}, for example, presents a concept design of using mobile AR to enhance the experience of live sports events.
Gamebot~\cite{DBLP:conf/chi/ZhiM20} uses a conversational interface to help users request data visualizations in watching NBA games. 
GameViews~\cite{DBLP:conf/chi/ZhiLSM19} uses simple visualizations (\emph{e.g.}, line charts)
to show in-game box scores of basketball games.
Omnioculars~\cite{lin2022the} uses interactive embedded visualizations to support in-game analysis of basketball games.
CourtVision~\cite{courtvision} is a commercial product that
allows inspectors to review basketball in-game data through simple, non-interactive embedded visualizations (e.g., text labels, progress bars).
Compared to traditional sports, 
most E-Sports already provide a personalized game viewing experience by default.
\cmo{Multiplayer Online Battle Arena} (MOBA) games, such as Defense of the Ancients2~\cite{dota} (Dota2) and League of Legends~\cite{lol} (LoL),
allow spectators to interact with the systems to inspect in-game data (\emph{e.g.}, points over time) of players or teams.
Nevertheless, these systems either display the data in separated panels 
or require viewers to explicitly interact with the system to request the data, 
inevitably distracting viewers from the game.
In contrast, we propose to use embedded visualizations and gaze interactions
to present extra data in game videos,
providing an intuitive, seamless, and engaging watching experience.

\subsection{Embedded Visualizations in Sports Videos}
Embedded visualizations have been widely used for sports data
due to their ability to show the data into its physical context (\emph{e.g.}, a basketball court).
Early works mainly embedded the data into static court diagrams.
Examples such as CourtVision~\cite{goldsberry2012courtvision} (basketball), 
StatCast Dashboard~\cite{DBLP:journals/cga/LageOCCDS16} (baseball),
and SnapShot~\cite{DBLP:journals/tvcg/PileggiSBS12} (ice hockey) 
display density maps on top of court diagrams to show sports events, 
such as successful shots. 
Recent progress in CV
now allows embedding visualizations directly into sports videos instead of just court diagrams.
For example, Stein et al.~\cite{DBLP:journals/tvcg/SteinJLBZGSAGK18, DBLP:conf/bdva/SteinBHSNSGKJ18} 
developed a method to automatically extract and visualize data from and in soccer videos.
Zhu-Tian et al.~\cite{VisCommentator, chen2022sporthesia} explored the design of augmented sports videos
and introduce fast prototyping tools to help users create augmented videos for racket-based sports
by using direct manipulation and textual comments.
However,
these works mainly target experts for analytic and authoring purposes.
More recently, 
researchers have started to explore embedded visualizations in live game-watching scenarios.
Yao et al.~\cite{yao2022visualization} proposed the notion of \emph{visualization in motion}
to depict visualizations that are moving relative to the viewer
and summarized a design space for it.
Lin et al.~\cite{lin2022the} presented a design framework 
for embedded visualizations to facilitate in-game analysis when watching basketball games.
Yet, all the above works only evaluated visualizations in simulated scenarios 
(e.g., moving charts on white backgrounds, 3D virtual sports games).
We design our embedded visualizations based on these prior works
but particularly target real basketball videos,
with the aim to understand how embedded visualizations can improve casual fans' game-watching experience.

\subsection{Computer Vision for Embedded Visualizations}
Recent years have shown remarkable advances in CV techniques based on deep learning.
Researchers have achieved unprecedented success in a broad range of tasks including
object detection~\cite{yolox2021, liu2021Swin}, 
object tracking~\cite{aharon2022bot, zhang2022bytetrack}, 
pose estimation~\cite{xu2022vitpose}, 
and segmentation~\cite{chen2022vitadapter}.
Thanks to this progress,
more and more data can be extracted from videos (e.g., \cite{DBLP:conf/cvpr/DeliegeCGSDNGMD21, DBLP:conf/cvpr/RamanathanHAG0F16, DBLP:journals/corr/KayCSZHVVGBNSZ17, DBLP:conf/cvpr/GuSRVPLVTRSSM18}),
opening new opportunities for sports analytics.
For example, 
the positions of the players and the ball~\cite{sportvu}, as well as other tracking data~\cite{nbashotcharts, nbastats}, 
of each NBA game are extracted and shared online.
We refer the reader to Shih~\cite{shih2017survey} for a comprehensive survey on content-aware video analysis for sports.
Furthermore, these new CV techniques ease the embedding
of visualizations into the video scenes,
which is recognized as a grand challenge in situated visualization~\cite{DBLP:conf/chi/EnsBCESWPABDDGH21}.
Embedding visualizations into sports videos
requires a CV pipeline to complete tasks such as
detecting, 
tracking, 
and segmenting 
the players from the video, 
estimating their pose,
calibrating the camera~\cite{DBLP:journals/pami/Zhang00},
and sometimes reconstructing the 3D scene~\cite{DBLP:journals/cacm/MildenhallSTBRN22}.
Prior works~\cite{VisCommentator, chen2022sporthesia, DBLP:journals/tvcg/SteinJLBZGSAGK18} applied a simplified CV pipeline 
to process racket-based sports videos,
in which the players are separated, and the camera is mostly static.
However, it is much more difficult to embed visualizations into team sports videos (especially basketball)
since players overlap heavily and the camera typically moves rapidly.
While commercial systems~\cite{courtvision} can achieve good embedding results,
\cmo{they require videos collected from multiple cameras~\cite{nba_tracking} to register the visualizations.
To the best of our knowledge, there is no existing CV solution 
that can embed visualizations into basketball videos based solely on broadcasting videos.
}
The lack of such a solution inevitably hinders the research of embedded visualizations in complex, dynamic scenarios, such as team sports.
In this work,
we contribute a CV pipeline that consists of open-sourced modular components
to process team sports videos for embedding visualizations.

\subsection{Applications of Gaze Interactions}
There is a long history of interest in leveraging gaze for interactions 
due to its efficiency, expressiveness, and applicability in hands-free scenarios~\cite{DBLP:journals/tochi/MengesKS19, majaranta2014eye, DBLP:journals/pervasive/BullingG10}.
Gaze interactions either \emph{explicitly} or \emph{implicitly} leverage the gaze to interact with digital content. 
We focus on implicit methods
and refer readers to a more comprehensive review~\cite{majaranta2014eye} for further reading.

Implicit gaze-based systems use gaze as an implicit input source, 
usually in combination with other input modalities,
to facilitate interactions~\cite{DBLP:conf/chi/FeitWTPKKM17, majaranta2014eye}.
Given that reliable eye trackers are now affordable enough to be integrated into desktop and laptop computers,
researchers have leveraged implicit gaze interactions to support a variety of applications,
such as content annotation~\cite{DBLP:conf/chi/ChengSSYD15, DBLP:conf/chi/TemplierBH16},
video editing~\cite{retargeting1, retargeting2, shotselection},
and remote collaborations~\cite{DBLP:conf/chi/KuttLHP19, gazechat}.
The most relevant to our endeavor 
are attempts at adapting viewing content based on users' gaze.
The gaze-contingent display~\cite{DBLP:journals/cbsn/DuchowskiCM04}, for example,
shows a higher resolution on the area the user is focusing on.
Other examples include adjusting the playback speed of lecture videos based on the user's gaze~\cite{DBLP:conf/chi/NguyenL16}, or a tourist guide that directs a user's gaze to highlighted features in a panorama and adapts the audio introductions accordingly~\cite{DBLP:conf/chi/KwokKSAR19}.
Kurzhals et al.~\cite{gazecaption} have proposed a gaze-adaptive system that dynamically adjusts video captions' placement to optimize the viewing experience.
We also aim to use gaze to adjust video content but focus on augmented sports videos.

In the visualization field, research related to gaze mainly focuses on visualizing gaze data~\cite{sotaVisGaze} 
and analyzing users' gaze in viewing visualizations~\cite{DBLP:journals/tvcg/BorkinBKBYBPO16, DBLP:conf/uist/BylinskiiKOAMPD17}.
Only a few works~\cite{DBLP:journals/cgf/OkoeAJ14, DBLP:conf/fmt/SilvaSSEF16, DBLP:conf/etra/SilvaSVSEF18, DBLP:conf/iui/ShaoSES17} have explored leveraging the gaze to interact with visualization systems.
Silva et al.~\cite{DBLP:conf/etra/SilvaBJRWRS19} give a systematic review on eye tracking for visual analytics systems and current challenges.
We draw on this line of research and, to the best of our knowledge, are the first to explore gaze-aware embedded visualizations to improve the sports-watching experience.
\section{Formative Study with Basketball Fans}
\label{sec:formative}

To understand the practices, pain points, and solutions of casual fans in watching basketball videos,
we conducted a formative study.


\subsection{Study Setup}

\subsubsection{\textbf{Participants}}
\noindent
We recruited participants using university mailing lists and forums 
and pre-screened participants 
based on their fandom level, game-watching frequency, and basketball knowledge. 
In total, we recruited 
8 casual fans (P1-P8; M=3, F=5; Age: 18 - 35),
who only knew \quot{basic rules of basketball} 
and watched \quot{1 - 10 games per year}.
To better identify the pain points specific to casual fans, 
we further recruited 8 die-hard fans (P9 - P16; M=8; Age: 18 - 55), 
who knew \quot{basketball tactics and pros and cons of specific players} 
and watched \quot{at least 1 game per week}. No female die-hard fan responded to us.
All participants had normal vision or wore contact lenses or glasses to correct to normal vision.

\begin{table}[h]
    \centering
    \caption{The two game videos in the formative study.}
    \begin{tabular}{c c c c  c c}
     & \textbf{Date} & \textbf{Teams} & \textbf{Quarter} & \textbf{Duration (m:ss)} \\ \midrule
    \textbf{G1} & 2015.12.25 & GSW vs. CLE$^\ast$ & 4  & 9:02\\ \hline
    \textbf{G2} & 2015.12.22 & OKC vs. LAC$^\ast$ & 4  & 9:30\\
   \bottomrule
    \end{tabular}
    \label{tab:study1_games}
    \begin{flushleft}\footnotesize
    $^\ast$ GSW (Golden State Warriors), CLE (Cleveland Cavaliers), OKC (Oklahoma City Thunder), and LAC (Los Angeles Clippers)
    \end{flushleft}
\end{table}

\begin{figure*}[h]
    \centering
    \includegraphics[width=0.9\textwidth]{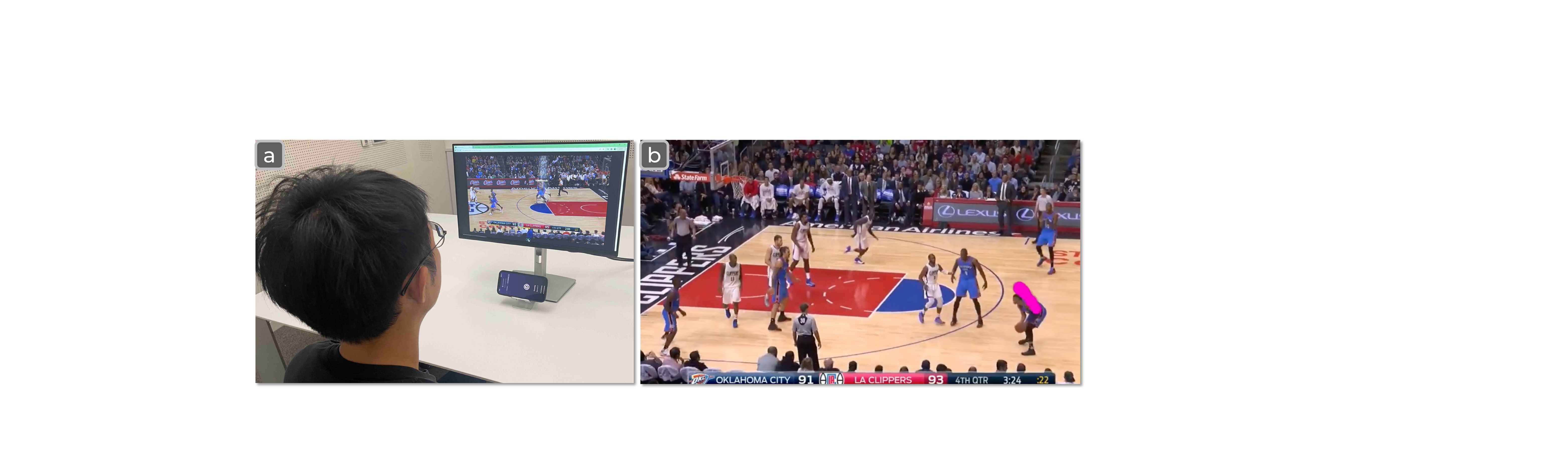}
    \caption{a) Formative study setup. 
    b) In the review phase, the participant's gaze is visualized and overlaid on the video.}
    \label{fig:study_setup}
\end{figure*}

\subsubsection{\textbf{Procedure}}
\noindent
We started each session 
by introducing our research motivation and study protocol.
The experimenter then conducted a semi-structured interview with each participant, 
focusing on their current practices, pain points, and solutions when watching live basketball games.
Next, we followed the format of \emph{contextual query}~\cite{hartson2012ux}
to ask participants to watch two videos (\autoref{tab:study1_games})
on a 24-inch monitor.
These two games were rated as top-30 games of the season 
and have been watched millions of times.
We collected think-aloud and gaze data during the game-watching process.
To collect the gaze data, we used Eyeware Beam~\cite{Eyeware},
a commercial software that leverages Apple's TrueDepth camera~\cite{TrueDepth}
to track the participant's head and gaze.
The participants sat approximately 60cm from the screen
and were asked to adjust the chair before watching the videos (\autoref{fig:study_setup}a).
The system was then calibrated and the participants
were allowed to move the head freely after the calibration.
We used a TrueDepth camera-based tracker as 
it provided sufficient accuracy~\cite{DBLP:conf/hci/GreinacherV20} 
for inspecting what video objects participants were looking at while watching the game, 
at a much lower budget.
For more fine-grained gaze data (e.g., saccades, fixation), more proficient eye-tracker would be required.

After watching each game video,
we asked participants to re-watch the game with their gaze data overlaid (\autoref{fig:study_setup}b)
and to elaborate on any confusion, data needs, insights, and excitement they had felt when watching the game for the first time.
Participants could pause the video in the review phase.
Each participant was compensated with a \$20 gift card for their time (1 hour).

\subsubsection{\textbf{Analysis}}
\begin{figure*}[h]
    \centering
    \includegraphics[width=0.99\textwidth]{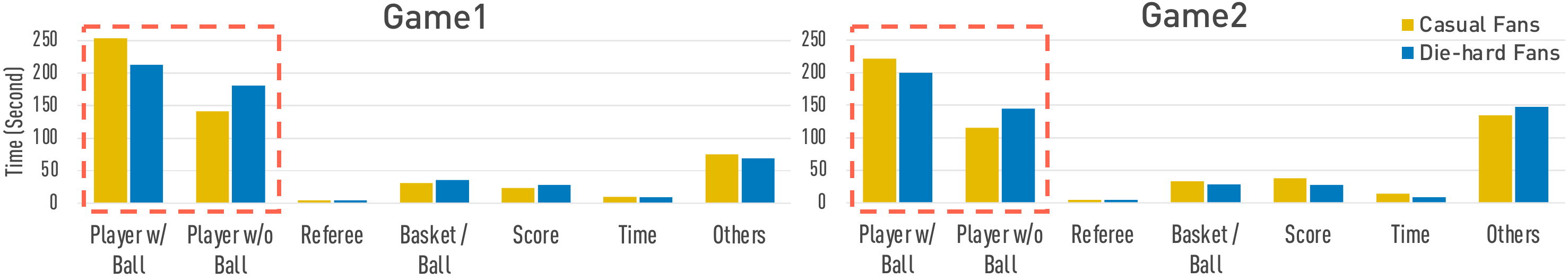}
    \caption{The gaze distribution in seconds shows that, compared to die-hard fans, casual fans spent more time watching the player with the ball in both games. Due to the small sample size, we discuss these results with a descriptive approach and focus more on other behavioral observations.}
    \label{fig:gaze_dist}
\end{figure*}

\noindent
Interviews and think-alouds were audio-recorded, 
transcribed, 
and analyzed using a reflexive thematic analysis~\cite{reflexiveTA}.
Three authors coded independently on the transcriptions to form sets of plausible codes
and iteratively refined the codes to converge on a single coding schema.
Besides, three authors analyzed the gaze data by manually annotating
the video objects each participant was looking at while watching the games.
The categories of objects (\autoref{fig:gaze_dist} x-axis)
were generated based on the data and prior knowledge.
We classified participants as looking at an object only when their gaze
rested on the object for at least 0.25 seconds (fixation duration~\cite{DBLP:conf/ozchi/PenkarLW12}). 
The duration when the gaze was moving to the object
was also annotated as looking at the object.

\subsection{Findings and Discussions}

All the casual fans
only watched \quot{important games, such as semi-finals or finals.} (P1)
They were neither familiar with basketball nor the NBA.
In comparison, 
the die-hard fans watched basketball games much more frequently.
They had a rich knowledge of basketball (e.g., tactics), 
knowing almost all NBA players and even their strengths and shortcomings. 
TV was the main way for all the participants to watch live basketball games.
Overall, for the casual fans, 
watching live basketball games was a leisure activity, such as hiking, 
but it was a more serious hobby for the die-hard fans. 

\subsubsection{\textbf{Casual Fans' Confusion in Watching Basketball Games}}

\noindent
In terms of the watching experience, 
all 16 participants confirmed that 
they were confused from time to time when watching basketball games
and that they would like to seek extra information, other than the data provided by the scoreboard and commentaries.
Some confusion is common among both casual and die-hard fans, 
such as questions like \quot{who got a foul?} and \quot{which team called the timeout?}
These questions can usually be resolved by \quot{watching the replay} (P15)
or simply by searching Google.
However, 
we did identify some confusing aspects specific to casual fans 
that cannot be easily resolved by the current methods 
and thus lead to a poor watching experience:

\paragraph{C1: \COne{}}
\label{para:C1}
When watching basketball games,
the casual fans often could not identify the important players
and felt that the players were just \quot{moving objects.} (P4)
The casual fans' inability to identify key players was also reflected in their gaze patterns.
In our study, 
we found that casual fans spent more time on the player with the ball than the die-hard fans (\autoref{fig:gaze_dist}),
since they
\quot{didn't notice other players' [off-ball] movement} (P1)
when watching the game.
As a result, 
the casual fans often missed important off-ball movements 
and felt that the ball \quot{magically fly to an open player.} (P1)
Moreover,
in some casual fans' gaze,
we noticed some rapid zigzag movement between the player with the ball and the other players, 
revealing their attempts to scan through the players.
P4, for example, explained that she was 
\quot{scan[ning] other players}
to predict the ball receiver at the next pass
while keeping an eye on the player with the ball,
leading to a heavy cognitive load.
In contrast,
the die-hard fans scanned through the players much more predictively
and often could directly identify the next ball receiver.

\paragraph{C2: \CTwo{}}
\label{para:C2}
The casual fans could hardly understand the in-game decisions 
of the players,
since the situated factors (e.g., players’ abilities) 
behind these decisions
were 
hard to interpret from the videos.
Consequently,
the casual fans could not appreciate the game at the same deep level as die-hard fans 
and had difficulties keeping pace with their experienced friends.
This was also revealed in the think-aloud data of the participants.
When watching the two videos,
the most frequent verbal comments from the casual fans were interjections, e.g., \quot{Oops}, \quot{Wooooow!}
Even in the follow-up review session, 
casual fans could hardly describe their thoughts while watching the games.
P3 acknowledged that 
she sometimes actually \quot{didn't totally understand} what was going on
but just felt excited. 
By contrast, the die-hard fans could clearly elaborate, comment on, and even suggest players' tactics when watching the games. 
Generally speaking, 
our study suggested that
the experience of watching games for
the casual fans was closer to ``feeling'' 
while the experience for the die-hard fans was closer to ``reading''. 

\paragraph{\cmo{C3: \CThree{}}}
\label{para:C3}
All the casual fans never searched the internet to seek data to resolve their confusion when watching the games.
\cmo{This was because the games were so fast and overwhelming
that they could miss key events when looking up websites.}
\cmo{Additionally,
the casual fans sometimes could not search for a player's data because they did not know the player's name.}
In contrast, 
the die-hard fans would search websites (e.g., ESPN) when watching the games,
though they also complained about the context switching between the games and the webpages.
According to the casual fans,
perhaps the best way to seek information about game understanding was to \quot{ask my [experienced] friends.} (P2)
Otherwise,
they would just \quot{let it [the confusions] go.}

\subsubsection{\textbf{Die-hard Fans' Suggestions for Understanding Basketball Games}}

\noindent
Since casual fans preferred to \quot{ask experienced friends} to seek information, 
we were interested in what information die-hard fans suggest for understanding a live basketball game.
Several critical insights were suggested by die-hard fans:

\paragraph{\cmo{Distinguishing between offense and defense.}}
Basketball, from a certain perspective, is a turn-based game.
A basketball game consists of multiple possessions (i.e., turns), 
in which the team that has possession of the ball is on offense, and the other team is on defense.
\cmo{
A player can have completely different roles, tactics, and behaviors between offense and defense.
Being aware of players' offense and defense status can help casual fans better understand and follow the game.
}

\paragraph{Identifying Key Players.}
While basketball is a team sport, 
the importance of each player, especially when she/he is on offense, is different.
Generally speaking, 
on the offensive side,
\emph{the player with the ball}
and \emph{the ball receiver} at the next pass are the most important ones.
\emph{Players with open spaces} are also critical to the offensive team as they have a higher chance of making the goal.
On the defensive side,
all the defenders guarding the player with the ball are important.
By identifying these key players, 
the die-hard fans could watch the game more effectively and predictably.
In addition to the aforementioned key players,
we also discussed other players with the die-hard fans,
such as offensive helpers who play screens.
Overall, they suggested not helping casual fans identify these players, 
as their contributions to the possession outcome (e.g., a goal) are not explicit 
and thus can confuse casual fans.

\paragraph{Understanding In-game Decisions.}
Knowing players' offensive and defensive abilities is essential to understanding their in-game decisions. 
The die-hard fans suggested two metrics
to help casual fans understand the players' abilities.
For offensive players, we can present their location-based \emph{expected point value},
which measures how many points a player is expected to make if they shoot at a specific location. 
For defensive players, 
we can present their location-based \emph{percentage points difference},
which measures how much the field goal percentage of a player changes when being defended by the defensive player.
Both metrics can be calculated or directly obtained by using the data from the Official NBA Stats website~\cite{nbastats}.  
The die-hard fans also suggested visualizing the one-on-one relationships 
between offensive and defensive players, 
which can reveal interactions between the players and their tactics (e.g., defensive switching).

\subsection{Summary}

{\renewcommand{\arraystretch}{1.2}
\begin{table}[h]
\small
    \centering
    \vspace{-1mm}
    \caption{Three design requirements for assisting casual fans in game watching derived from the formative study.}
    \begin{tabular}{p{0.4\linewidth} p{0.4\linewidth}}
    \toprule
    
    \cmo{\normalsize\textbf{Findings}}  & \cmo{\normalsize\textbf{Design Requirements}}  \\ \midrule
    \cmo{\emph{C1}: \COne{}}  & \cmo{R1: Guide the user's attention to the important offensive and defensive players. (\autoref{sec:direct_att})} \\ \hline
    
    \cmo{\emph{C2}: \CTwo{}} & \cmo{R2: Visualize players' offensive and defensive abilities. (\autoref{sec:vis_abilities})} \\ \hline
    
    \cmo{\emph{C3}: \CThree{}} & \cmo{R3: Provide a fast and seamless method to retrieve data of interest.
    (\autoref{sec:gaze_int})} \\ \bottomrule

    \end{tabular}
    \vspace{-4mm}
    \label{tab:findings}
\end{table}

}

\noindent
In summary,
the casual fans were often confused about the key players and their in-game decisions,
but rarely sought data to resolve their confusion because
the searching process is slow and distracting.
To help casual fans better understand the game, 
the die-hard fans suggested a few critical insights, including
\cmo{distinguishing between offense and defense},
identifying key players, 
and understanding players' in-game decisions.
\cmo{By interpreting these findings and suggestions,
we derived three design requirements (\autoref{tab:findings})
and designed \system{}.}
Next, we first introduce 
a CV pipeline (\autoref{sec:cv_pipeline}) to enable \system{},
followed by a set of gaze-moderated embedded visualizations 
(\autoref{sec:gaze_moderated_vis}).

\section{A CV Pipeline for Embedding Visualizations} 
\label{sec:cv_pipeline}

\begin{figure*}[hb]
    \centering
    \includegraphics[width=0.9\textwidth]{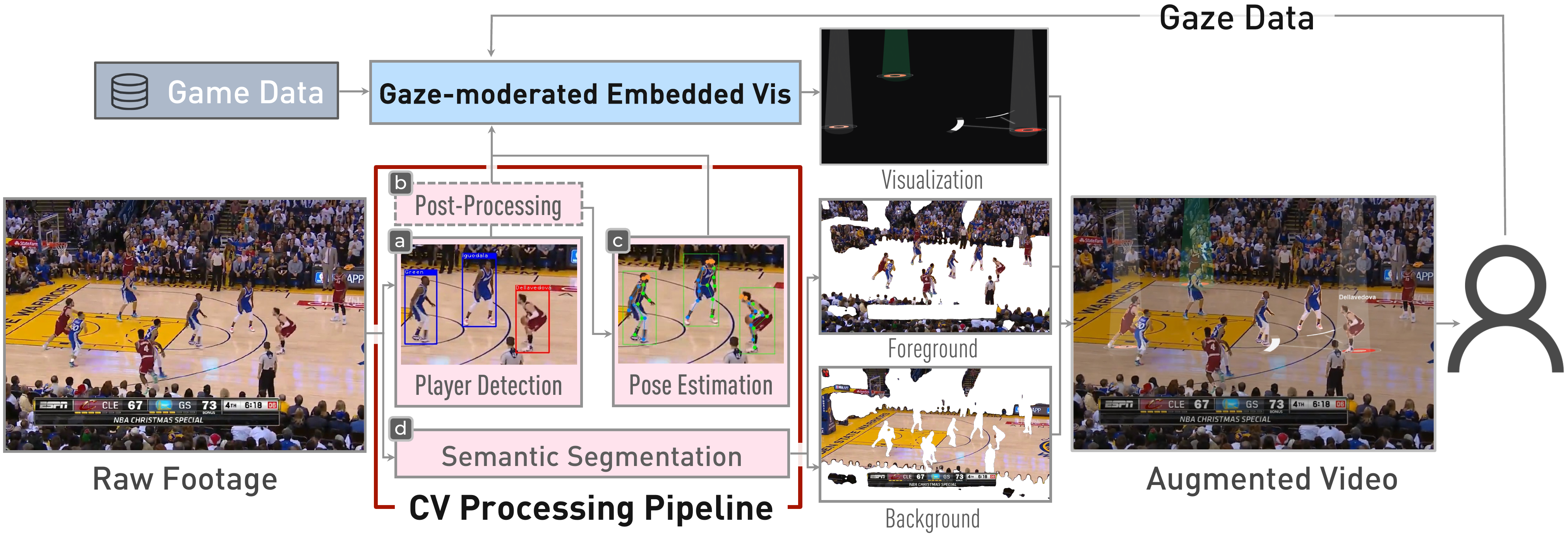}
    \caption{Our CV pipeline takes a raw video as the input, 
    outputs the bounding box, identity, and key points of each player, 
    and separates the image frame into the foreground (humans) and background (all others). 
    The bounding boxes, identities, and key points are used to create visualizations,
    which are then composited with the foreground and background to form the augmented video.}
    \label{fig:cv_pipeline}
\end{figure*}

\noindent
To embed visualizations into a basketball video, 
we need to recognize the players (e.g., bounding box, identity, and key points)
and segment them from the background.
To this end,
we designed a CV pipeline (\autoref{fig:cv_pipeline})
\cmo{to pre-process team sports videos.}

\subsection{Recognizing the Players}
To embed visualizations for a player,
one must first recognize the player in the video.
For example, to display a label with the name of a player, 
the system needs to detect the video object that corresponds to the player (bounding box and identity)
and the player's key body joints (key points) for placing the label.
Given a raw video frame,
we obtain this information for each player
via three steps:

\paragraph{\textbf{Step 1. Player Detection} (\autoref{fig:cv_pipeline}a).}
\cmo{
To obtain the players' bounding boxes and identities in a video frame,
we use an object detection model to locate and classify each player into different categories.
Different from common object detection tasks,
we use the players' identities as their categories.
In our implementation,
we fine-tuned a COCO~\cite{caesar2018coco} pretrained YoLoX~\cite{ge2021yolox} model
on an NBA player dataset (details in Appendix A). 
The output of the model is
a set of bounding boxes associated with their identities and confidence scores (i.e., $score_{c}$).
By convention,
only those bounding boxes with $score_{c}$ greater than a threshold $T_{high}$
are considered successful detections.
}

\paragraph{\textbf{\cmo{Step 2. Post-Processing}} (\autoref{fig:cv_pipeline}b).}
\cmo{
One limitation of the object detector is that
it only utilizes the players' visual appearance information
to determine their confidence score.
Consequently,
the detector can assign low confidence scores to players 
whose visual qualities are low (e.g., when they are occluded by others)
and filter them out.
We thus use object trackers to exploit the players' motion information
to complement the detector.
An object tracker stores the history of an object's bounding boxes in the previous frames
and can predict the object's bounding box in the next frame by using a Kalman filter.
We use object trackers as follows:}

\begin{enumerate}[nosep]
    \item 
\cmo{
    For a frame $F_t$,
    we divide all the detected bounding boxes into three clusters based on their $score_{c}$: 
    high-quality boxes ($score_{c} > T_{high}$),
    low-quality boxes ($T_{low} < score_{c} < T_{high}$),
    and rejected boxes ($score_{c} < T_{low}$).
    }
    
    \item 
    \cmo{
    For each high-quality box, 
    we match it with the trackers in the previous frame $F_{t-1}$
    by calculating the Intersection over Union (IoU)
    between the box and the predicted boxes of the trackers.
    A tracker is considered as \emph{matched} with the high-quality box if it maximizes the IoU.
    If matching is successful, 
    we assign the matched tracker to the box;
    otherwise, we initialize a new object tracker for the  box.
    }
    
    \item 
    \cmo{
    For each low-quality box, 
    we match it with the remaining trackers 
    (i.e., those that have not been matched with any high-quality boxes).
    If matching is successful, 
    we assign the matched tracker to the box.
    }

    \item 
    \cmo{Finally, we output all the boxes with matched trackers.}

\end{enumerate}

\noindent
\cmo{
Intuitively, 
this method uses the motion information of the players to select 
some low-quality bounding boxes to complement the output of the detector.
We refer the readers to Bot-SORT~\cite{aharon2022bot} for more details
about the tracker and the matching process.
}

\paragraph{\textbf{Step 3. Pose Estimation} (\autoref{fig:cv_pipeline}c).}
We use a pose estimation model to obtain the players' key points, 
such as head, hands, hip, and feet.
In our implementation,
we first used the bounding boxes produced in Step. 2
to extract the players from the video frame
and then fed those boxes to ViTPose~\cite{xu2022vitpose} 
to estimate the key points. 

\subsection{Separating Foreground and Background}
\cmo{
According to previous works~\cite{VisCommentator, lin2022the},
embedded visualizations for sports, such as empty areas,
are often placed on the ground, beneath the players' feet.
}
To achieve this, 
we need to separate the video frame into the foreground (the objects \cmo{on the ground}) and background (\cmo{the ground}),
draw the visualizations onto the background,
and finally \cmo{overlay} the foreground on the background to
form an augmented video frame (\autoref{fig:cv_pipeline}d).
\cmo{Ideally, all the objects should be segmented from the ground.
To simplify the segmentation process, 
we decided to only segment humans from the video as the foreground
and leave the remaining pixels as the background,
as humans are the major objects on the ground in a basketball video.
}
In our implementation, 
we used a ViT-Adapter~\cite{chen2022vitadapter} trained on COCO 164K~\cite{caesar2018coco}
to perform binary semantic segmentation to segment the humans.

\subsection{\cmo{Computational Evaluation}}
\cmo{
To evaluate the performance of our pipeline, we conducted several experiments focused on three main questions: 
1) Can the object detector detect the players?;
2) Can the post-processing step improve the detections?;
3) How much time does each step take?
To answer these three questions, 
we manually annotated the bounding box and identity of each player 
in each frame of the two game videos used in~\autoref{sec:formative} (i.e., G1 and G2).
We then split each video into clips and allocated 70\% for training and 30\% for testing.
To accelerate the training process, 
we sampled every tenth frame from the training clips and used only these frames for training.
This is because the consecutive frames often contain redundant information.
Despite this, the testing was conducted on all frames in the testing clips.
The details of the dataset can be found in Appendix A.
We trained and evaluated the detector on G1 and G2 separately, 
using their default hyperparameters whenever possible.
}

\cmo{
We did not evaluate the accuracy of the Pose Estimation and Semantic Segmentation steps
because we used off-the-shelf models for their standard tasks without any fine-tuning in these two steps.
Yet, their performance for basketball videos can be qualitatively evaluated by inspecting the augmented videos
provided in the supplemental material.
}

\begin{table}[h]
 \centering
 \small
    \caption{\cmo{Average Precision of the Player Detection and Post-processing steps.}}
    \label{tab:model_exps}
    {\color{cb_blue}
        \begin{tabular}{l l l l l}
        \toprule
        \textbf{Dataset}   & \textbf{Step}  & $\mathbf{AP_{50:95}}$ & $\mathbf{AP_{50}}$ & $\mathbf{AP_{75}}$  \\ \midrule
        \multirow{2}{*}{\textbf{G1}}& Player Detection & 65.4 & 83.6 & 76.2\\
                                         &  Post-Processing & \textbf{69.2} (+3.8) & \textbf{87.9} (+4.3) & \textbf{79.4} (+3.2) \\ \hline
    
        \multirow{2}{*}{\cmo{\textbf{G2}}}& Player Detection & 70.7 & 86.1 & 82.3\\
                                         &  Post-Processing & \textbf{75.0} (+4.3) & \textbf{90.0} (+3.9) & \textbf{85.3} (+3.0) \\ \hline
                    \textbf{COCO}$^\ast$ & YoLoX & 51.2 & 69.6 & 55.7 \\
        \bottomrule
        \end{tabular}
    }

    \begin{flushleft}\footnotesize
    $^\ast$Due to the lack of benchmarks, we provide YoloX's performances on COCO as a reference. However, it does not serve as a comparative baseline.
    \end{flushleft}
    \vspace{-3mm}
\end{table}

\cmo{
Table.~\ref{tab:model_exps} 
shows the performance of our fine-tuned object detector on the testing clips of G1 and G2.
To access the detector,
we followed the convention to calculate the Average Precision (AP) metrics over different IoUs.
The higher the AP, the better it is.
$AP_{50:95}$ is the average AP over different IoU, from 0.5 to 0.95 with step 0.05.
$AP_{50}$ and $AP_{75}$ are the APs calculated at IoU 0.5 and 0.75, respectively.
The larger the IoU, the stricter the metric will be.
Overall, the results reveal that our fine-tuned object detector 
can perform well in detecting players.
}
\cmo{
Furthermore, 
all the APs increase
after applying the post-processing step, which shows
that the post-processing step is useful and can complement the detector to improve its results.
}

\begin{table}[h]
 \centering
 \small
    \caption{\cmo{Time cost of each step.}}
    \captionsetup{width=.8\textwidth}
    \label{tab:model_time}
    {\color{cb_blue}
        \begin{tabular}{l r}
        \toprule
        \textbf{Step}  & \textbf{Time (ms)} \\ \midrule
        Player Detection & 31.98 \\
        Post-Processing & 2.40 \\
        Pose Estimation & 121.00 \\ \hline
        Semantic Seg.$^\ast$ & 3674.96 \\  \bottomrule
        \end{tabular}
    }
    \begin{center}\footnotesize
        $^\ast$Semantic Segmentation can run in parallel with other steps.
    \end{center}
    \vspace{-3mm}
\end{table}

\cmo{
In terms of time performance,
Table.~\ref{tab:model_time} shows the average time in milliseconds (ms) each step takes to process a video frame.
We tested the pipeline on a machine with a Nvidia Tesla V100 graphic card
and only counted the inference time of the models 
by excluding the model and dataset loading time.
Overall, the Player Detection and Post-processing steps use ~34ms for one frame,
almost achieving 30FPS.
Other steps, especially the Semantic Segmentation step, need longer to process one frame.
These results show that the semantic segmentation model we used is the bottleneck
for extending the pipeline to support real-time scenarios.
}


\subsection{\cmo{Extendibility}, Generalizability, and Limitations}
\label{sec:cv_limitation}

The contribution of our pipeline does not lie in the individual components
but a workable solution that shows
which CV models are required
and how they can be composited together 
to process basketball videos for the purpose of embedding visualizations into videos.
To inspire future research,
we further discuss the \cmo{extendibility}, generalizability and limitations of the pipeline:

\vspace{-0.5mm}
\paragraph{\cmo{Extendibility.}}
\cmo{
Our pipeline can be extended for better performance.
To improve the accuracy,
we can try using better models or adding extra components to the post-processing step to improve the detections.
For example,
in our implementation, 
we further interpolated and smoothed the bounding boxes for the user study.
To improve the efficiency,
we can use faster models, more powerful graphic cards, or remove the Semantic Segmentation step if visualizations on the ground are not needed.
Overall, our pipeline can serve as a reference for other researchers to develop their own systems for their specific scenarios, videos, and tasks.
}

\vspace{-0.5mm}
\paragraph{Generalizability.}
The CV pipeline 
can be applied to other basketball videos
and even other team sports videos.
For example, there are about 450 players in the NBA~\cite{NBA}.
To generalize the pipeline to \cmo{other NBA game videos},
we need to develop a player dataset of these 450 players
to fine-tune the detector.
\cmo{Note that it is not necessary to develop a player dataset for each video.
Our experiments showed that the detector could detect players on unseen testing clips even if it was trained only on the training clips.
If the player dataset is large enough, 
the detector fine-tuned on it can be applied to any NBA game video.
}
This is not impossible
as modern deep learning-based image classifiers can
achieve superhuman performance on tasks with more than 1000 classes~\cite{yu2022coca}
and many priors can be used to optimize the model results, 
e.g., there are no more than 24 players in a game.

\vspace{-0.5mm}
\paragraph{Limitations.}
\cmo{The pipeline and the evaluation have a few limitations.
First, 
as shown in Table.~\ref{tab:model_time}, 
the processing time of our pipeline
for one frame is about 4 seconds.
While the Semantic Segmentation step can run in parallel with others,
our implementation can only pre-process the game videos instead of running in real-time.}
Second, 
our pipeline only extracts 2D information from the video,
limiting the design space of available embedded visualizations.
For example, without the camera parameters, we cannot display visualizations that are static relative to the ground, such as trajectories.
\cmo{
In reality, the camera parameters can be provided by the producer of the video
or estimated using camera calibration techniques (e.g., \cite{Sha_2020_CVPR, DBLP:journals/tmm/HuCWC11}).
In our implementation,
similar to prior research~\cite{chen2022sporthesia, VisCommentator},
we treated the camera parameters as partially known meta information} 
to display visualizations that are static relative to the players.
%
%
\cmo{
Third, we only evaluated the pipeline on G1 and G2.
A larger video dataset with more ground truth labels
is required
to fully test the pipeline.
We consider developing
such a sports video dataset 
beyond the scope of this work and leave it for the future.
}
\section{Gaze-moderated Embedded Visualizations}
\label{sec:gaze_moderated_vis}

\begin{figure*}[hb]
    \centering
    \includegraphics[width=0.9\textwidth]{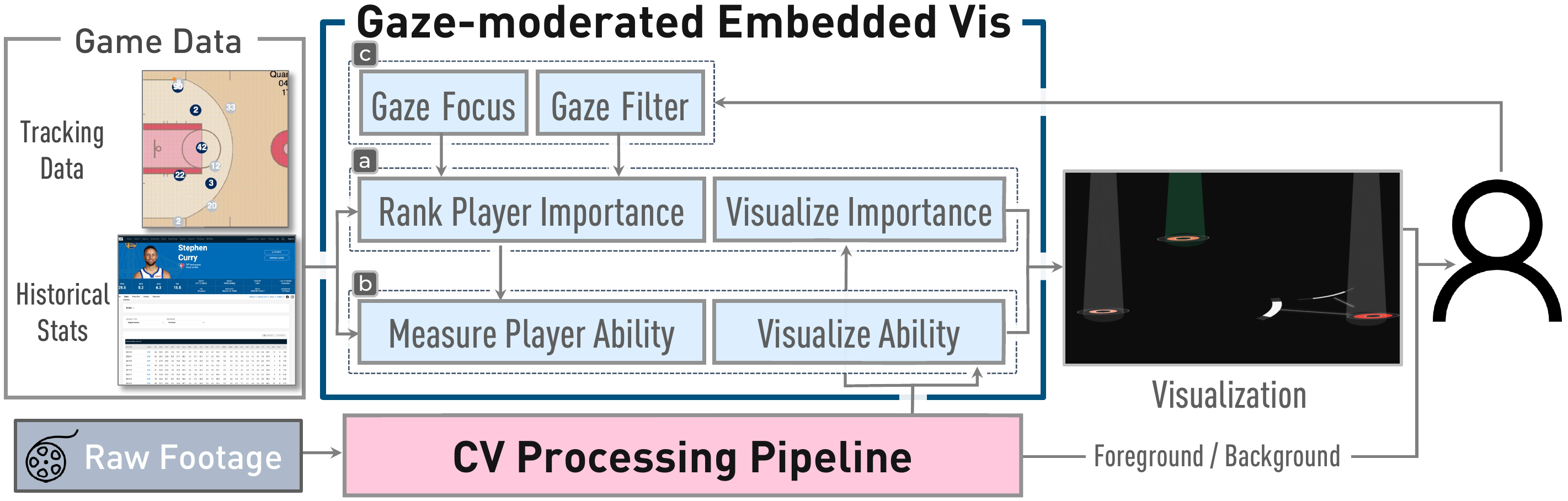}
    \caption{
        The system takes positional tracking data and historical stats as input to calculate the players' importance and offensive and defensive abilities.
        Only the important players and their offensive and defensive abilities will be highlighted and visualized in the video.
        The user can use gaze points to adjust the players' importance levels, as well as controlling whose abilities to show.
    }
    \label{fig:gaze_vis}
\end{figure*}

\noindent
Based on the identified design requirements from the formative study, 
we designed a set of gaze interactions that can naturally guide and respond to the user's attention through gaze tracking without explicit user input.
Our gaze-moderated embedded visualizations 
\cmo{
1) guide audiences' attention 
and 2) reveal players' offensive and defensive abilities} 
and 3) update the embedded visualizations based on gaze. 
\cmo{The system flow is shown in \autoref{fig:gaze_vis}a-c.}

\subsection{Guiding Audiences' Attention}
\label{sec:direct_att}

To \cmo{help casual fans identify the important players (\hyperref[para:C1]{R1})},
we first ranked the players' importance levels according to die-hard fans' suggestions
and then highlighted the players accordingly.

\subsubsection{\textbf{Ranking Players' Importance Levels}}

\noindent
Based on the formative study,
we adopted an offensive-first method to
rank the players' importance into three levels:

\begin{itemize}[topsep=0pt]
    \item [\textbf{Lv3 -}] \textbf{Key offensive players}: 
        The player with the ball, 
        the next ball receiver, 
        and the players with open spaces
        are considered as the most important offensive players.
        \cmo{
        When pre-processing the game videos,
        we used positional tracking data~\cite{sportvu}
        to identify which players had the ball or were with open spaces in each frame.
        Meanwhile, we looked ahead 1.8 seconds (selected empirically) 
        to find the next ball receiver.
        }
        \cmo{
        To extend our system to livestream scenarios in the future, 
        potential approaches could be to use machine learning models~\cite{DBLP:conf/soict/SanoN19} or the buffer time in video streaming to detect the next ball receiver.
        }        

    \item [\textbf{Lv2 -}] \textbf{Key defensive players}: 
        \cmo{
            The players who are defending the player with the ball are considered as the important defenders.
            In our implementation,
            inspired by previous work~\cite{tian2019use},
            we detect important defenders
            by checking which defenders were closest to the player with the ball within a time interval.
        }
        
    \item [\textbf{Lv1 -}] \textbf{Other players}:
    All other players who do not belong to \textbf{Lv3} and \textbf{Lv2} fall into this level.
\end{itemize}

\noindent
To detect if a player is in offense or defense,
we also used the positions of the players and the ball.
If a player or one of her/his teammates is the closest player to the ball within a predefined time interval (0.5s in our implementation),
she/he is in offense; otherwise, in defense.
Note that we deliberately ignored some important players,
such as those who play screens or specific tactics,
since casual fans usually cannot understand why these players are important.

\begin{figure*}[h]
    \centering
    \includegraphics[width=0.9\textwidth]{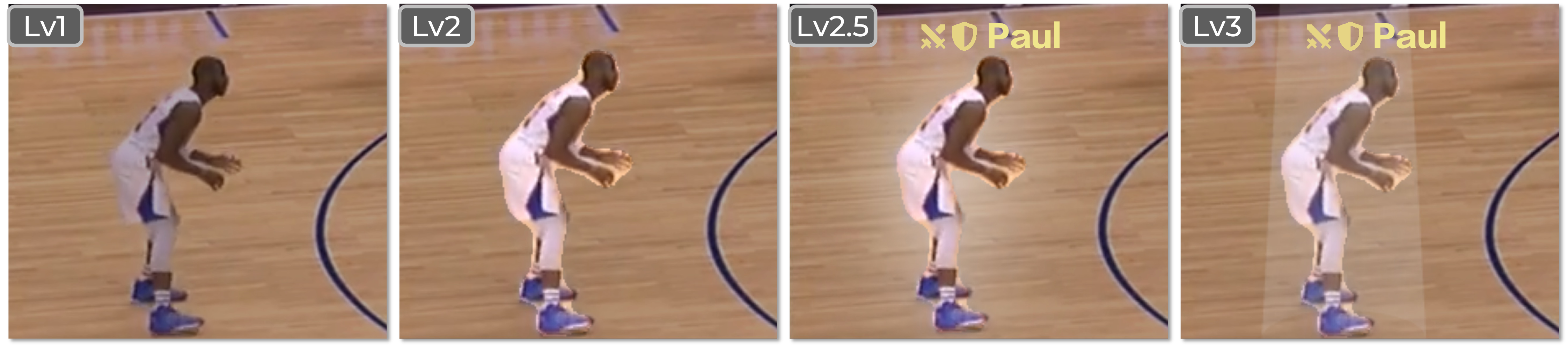}
    \caption{Visualization of various importance levels: 
        \textbf{Lv3}, key offensive players, \cmo{highlighted by a sportlight};
        \textbf{Lv2.5}, players of interest to the user, triggered by \GFo{}, \cmo{highlighted by a glowing effect} (\cmo{will be introduced in Sec.~\ref{sec:gaze_int}});
        \textbf{Lv2}, key defensive players, \cmo{highlighted by extra brightness};
        \textbf{Lv1}, other players, \cmo{no highlighting}.
    }
    \label{fig:att_lv}
\end{figure*}

\subsubsection{\textbf{Visualizing Importance Levels}}

\noindent
We designed multiple highlight effects to guide user attention to players at different importance levels (\autoref{fig:att_lv}).
We considered the effectiveness and aesthetics of the visualizations and iterate our designs to
make them intuitive and distinguishable, i.e., the one for a higher importance level is more attractive.
We displayed the name of players with importance levels greater than \textbf{Lv2}.
We also colored the name of ``star'' players in gold
with an icon showing their roles (i.e., \inlinefig{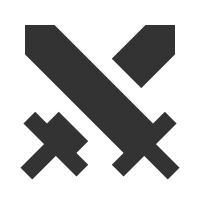} for good shooter and \inlinefig{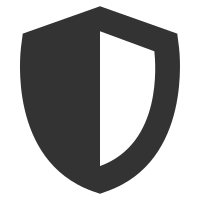} for good defender).
Furthermore, Lv3 spotlight encodes the different offense roles with color, \emph{green} for players with open space and \emph{white} for other key offensive players.

\subsection{Revealing Players' Abilities} 
\label{sec:vis_abilities}

\noindent
To help casual fans understand the players' abilities (\hyperref[para:C2]{R2}),
we computed and visualized two location-based metrics of the players whose attention level is higher than \textbf{Lv1} (\autoref{fig:gaze_vis}b).

\subsubsection{\textbf{Measuring Players' Offensive and Defensive Abilities}}

\noindent
We used two well-established metrics to indicate the players' offensive and defensive abilities:

\begin{itemize}[topsep=0pt]
    \item \textbf{Offense - Expected Point Value (EPV)} 
    measures how many points a player is expected to make if he/she shoots from the current position.
    In basketball, it is a value between 0 and 3.
    Fundamentally, the goal of offensive tactics in basketball games
    is to maximize the EPV of the shooter.
    Thus, visualizing the EPV can
    help casual fans better understand and evaluate the in-game decisions of offensive players (e.g., pass or shoot).
    We obtained the EPV for each player based on their historical shot records.
    Specifically, 
    we created a hexbin shot chart~\cite{shotchart} for each player 
    based on their historical shot records,
    in which the bins are grouped
    based on the shooting regions (defined by Official NBA Stats~\cite{nbastats}).
    We then
    calculated the EPV per region by multiplying the player's field goal percentage and points they can make in the region.
    The results were cached as an EPV map for efficient access in each frame.
    Figure~\ref{fig:epv_map} shows an example EPV map.

\begin{figure}[h]
    \centering
    \includegraphics[width=0.9\columnwidth]{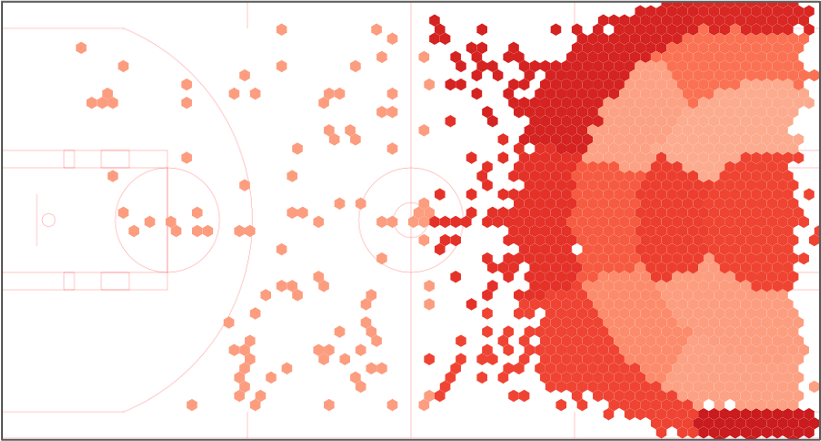}
    \caption{An EPV map of Stephen Curry based on his shooting records in the 2015-16 season. A darker color indicates a higher EPV.}
    \label{fig:epv_map}
    \vspace{-2mm}
\end{figure}

    \item \textbf{Defense - Percentage Points Difference (DIFF\%)} 
    \cmo{is a measure of a defender's ability to affect a shooter's shot percentage.}
    Good defenders will have a negative DIFF\% since they hold their opponent to a lower percentage than normal.
    \cmo{For example, Stephen Curry's DIFF\% is $-3.6$\%, which means on average, a shooter's shot percentage will decease by 3.6\% when being guarded by Curry.
    We acquired DIFF\% by regions for each player directly from NBA Stats~\cite{nbastats}.
    }
    Besides DIFF\%, 
    the distance between a defender and the offensive player with the ball (DIST) is critical to the defensive performance.
    We calculated DIST based on the positional tracking data.
\end{itemize}

\begin{figure*}[h]
    \centering
    \includegraphics[width=0.98\textwidth]{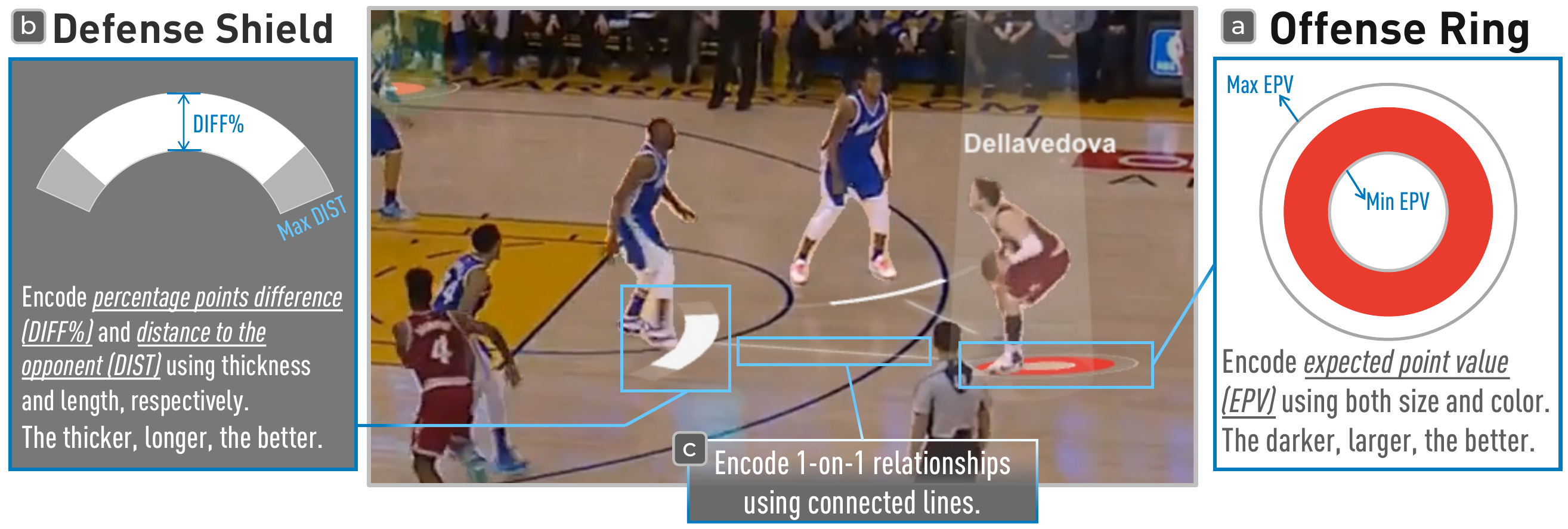}
    \caption{Three embedded visualizations for in-game data:
        a) \OffR{} shows the offensive performance of an offensive player. The darker, larger, the better.
        b) \DefS{} shows the defensive performance of the defender. The thicker, longer, the better.
        c) \OOOL{} shows the one-on-one relationship between the offensive player with the ball and the defenders.
    }
    \label{fig:od_data}
    \vspace{-3mm}
\end{figure*}

\subsubsection{\textbf{Visualizing Players' Offensive and Defensive Abilities}}

\noindent
Grounded in the design space proposed by prior work~\cite{lin2022the},
we designed three embedded visualizations to present the offensive (EPV) metric,
defensive (DIFF\% and DIST) metric,
and the one-on-one relationship between the defenders and the offensive player with the ball:

\begin{itemize}[topsep=0pt]
    \item \textbf{\OffR{}} (\autoref{fig:od_data}a) presents the player's 
    location-based EPV, 
    where a larger ring with darker color indicates better offensive ability.  
    The inner and outer rings represent the minimum and maximum of possible EPV (i.e., 0 and 3).
    We used both the size and color of the middle ring to encode the player's EPV at the current position for easier interpretation.

    \item \textbf{\DefS{}} (\autoref{fig:od_data}b) 
    represents the defender's location-based DIFF\% and DIST in an arc shape,
    where a thicker and longer arc indicates better defensive ability. 
    The thickness of the ``shield'' encodes the inverse DIFF\% (a negative value) to make the visualization intuitive. 
    The arc length of the ``shield'' encodes the subtraction of DIST from \emph{maximum guarding distance}, 
    since a larger DIST indicates lower pressure from the defender to the player with the ball.
    We displayed an outer border of the ``shield'' to show the maximal guarding distance
    (empirically selected as 12 feet) for comparison.

    \item \textbf{\OOOL{}} (\autoref{fig:od_data}c) 
    visualize the one-on-one relationship between the key defenders and the offensive player with the ball.
    The player with the ball can be defended by multiple defenders.
\end{itemize}

\noindent
All these three visualizations are updated dynamically in the game based on the players' positions.
We also darkened the background image to provide enough contrast for reading the visualizations.

\begin{figure*}[hb]
    \centering
    \includegraphics[width=0.9\textwidth]{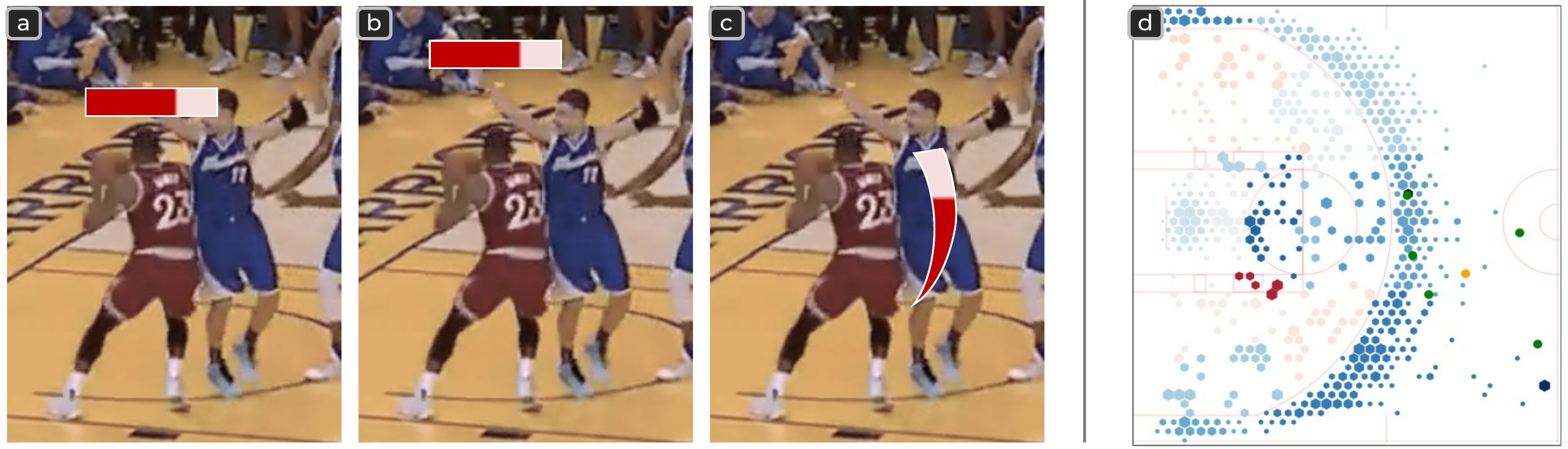}
    \caption{\cmo{
    Left: Three design alternatives for \OffR{}.
    a) Displaying the data on top of the player can occlude other players.
    b) Moving the visualization higher (e.g., the design in CourtVision~\cite{courtvision}) can make it hard to connect to the target player.
    c) Displaying the data aside of the players (e.g., the shot meter in NBA 2K~\cite{nba2k}) can also occlude other players.
    Right: An experimental EPV map of Steven Curry encodes his shooting frequency and EPV
    by using the size and divergent color scale. Different from \autoref{fig:epv_map}, the bins in this EPV map are not grouped by regions.
    }
    }
    \label{fig:alternatives}
\end{figure*}

\subsubsection{\cmo{\textbf{Design Process and Alternatives}}}

\noindent
\cmo{
We finalized our designs through multiple rounds of iterations, especially for \OffR{}.
Two considerations mainly drove our decision to use a ring placed on the ground ---
the visualization should 1) tightly connect with the target player
and 2) avoid occluding other objects.
Similar designs were used in previous research~\cite{lin2022the} and basketball video games~\cite{nba2k}.
Figure.~\ref{fig:alternatives}a-c show some alternative designs we explored but none of them were satisfactory.}

\cmo{
When designing the visual encoding of \OffR{},
we first used the size of the ring to encode a player's shooting frequency 
and a divergent color scale to encode the player's EPV,
with the league average EPV as the midpoint.
Figure.~\ref{fig:alternatives}d shows an EPV map we created based on this encoding schema.
However, in a pilot study, 
we found that this encoding schema was too complex to interpret for casual fans.
For example, 
when the size of the ring is small (low shooting frequency) but the color is dark blue (high EPV), 
casual fans cannot judge if this is a good chance for the player to shoot or not.
Thus, we decided to remove the encoding of shooting frequency 
and use both size and color to encode EPV.
However, this could still be confusing 
as the size scale is sequential but the color scale is divergent.
Consequently, we decided to use a sequential color scale instead of a conventional divergent one.
This design was found to be easy to understand with clear messages 
(i.e., the bigger and darker, the better) 
to improve game understanding for casual fans.
Different design decisions could be made for other purposes or fans, e.g., for analytic purposes or die-hard fans~\cite{lin2022the}.
}

\subsection{Gaze-based Interactions}
\label{sec:gaze_int}

\begin{figure*}[hbt]
    \centering
    \includegraphics[width=1\textwidth]{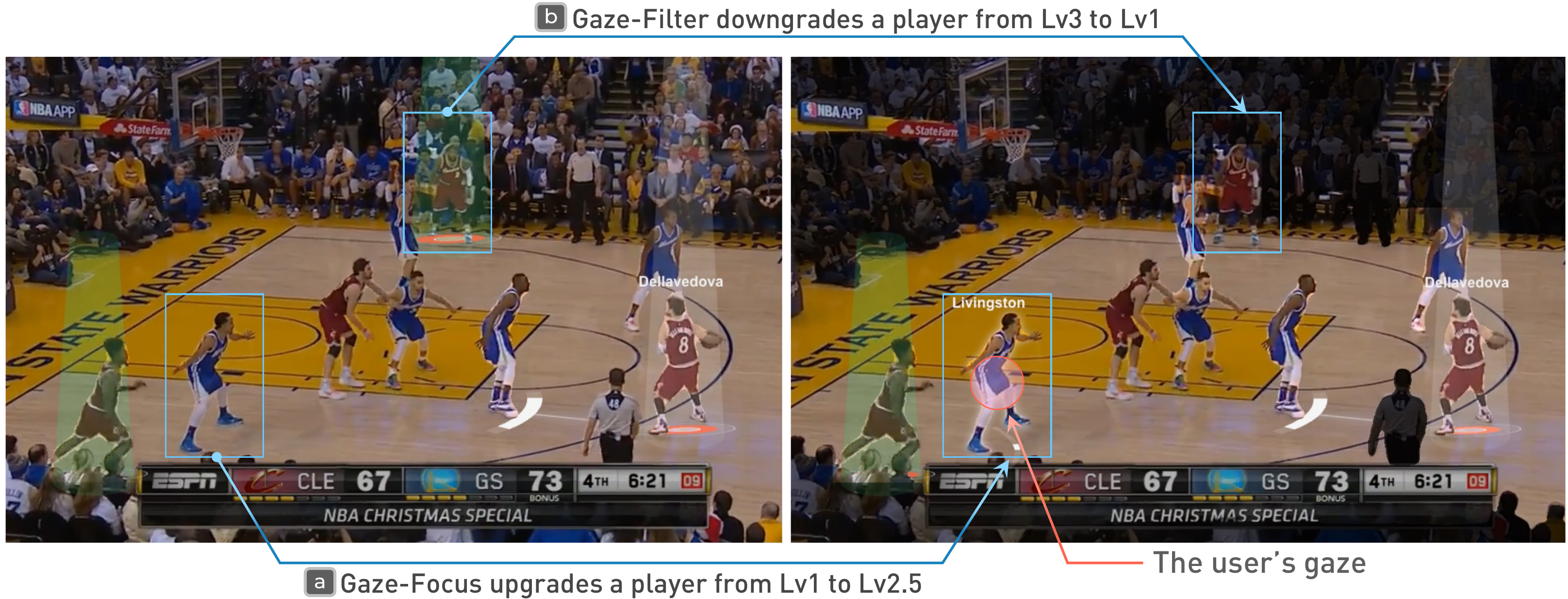}
    \vspace{-3mm}
    \caption{Two gaze interactions to adjust the embedded visualizations:
        a) \GFo{} lifts the importance level and shows in-game data of players that are of interest to the user.
        b) \GFi{} drops the importance level of video objects who are not the user's focus (e.g., open players and audiences out of focus).
    }
    \vspace{-2mm}
    \label{fig:gaze_int}
\end{figure*}



\noindent
To help casual fans seamlessly and efficiently access data of players they are interested in
while watching the game (\hyperref[para:C3]{R3}),
we \cmo{explored using gaze as an input signal} and designed two gaze interactions (\autoref{fig:gaze_vis}c).
\cmo{The fans can still be guided by the visualizations to identify important offensive players.}


\subsubsection{\textbf{\GFo{} -- Fetching Data of Players of Interest to the User}}
\noindent
\emph{\GFo{}} 
allows the user to express their interests in players through gaze dwelling, which lifts the importance level of the players.
\GFo{} comprises the following three considerations:
\begin{itemize}[topsep=0pt]
    \item \textbf{Trigger}: 
    According to our formative study, the users' gaze can move rapidly between, and across, players.
    To avoid showing data of players glanced over by the user accidentally, 
    we defined a ``dwell time''~\cite{DBLP:conf/ozchi/PenkarLW12} for the interaction.
    The user needs to dwell her/his gaze on a player for \cmo{0.25} seconds to trigger the interaction.

    \item \textbf{Visual feedback}:
    To help the user realize that she/he is gazing at a player
    and triggering the interaction,
    we designed a highlight effect (\autoref{fig:att_lv}c) in which 
    the glow of the player 
    will gradually increase when the user is gazing at the player, until the interaction is triggered.
    This design provides continuous visual feedback for the user 
    while conforming with the visual design of importance levels. 

    \item \textbf{Outcome}: 
    \cmo{Once the user triggers the interaction, \system{} lifts the targeted player to \textbf{Lv2.5} if she/he is currently at a lower level (\autoref{fig:gaze_int}a).}
    As a result, the system will also visualize the name and offensive or defensive data of the player.
    Lastly, when the user moves her/his gaze away from the player, 
    the player will stay in \textbf{Lv2.5} for 1.8s (selected empirically) before reverting to their original player importance level.
    We designed such a lasting duration to cope with the users' rapid saccade in game watching.
\end{itemize}

\subsubsection{\textbf{\GFi{} -- De-emphasizing Video Objects Out of the Sight}}
\noindent
To prevent users from being overwhelmed by too many \textbf{Lv3} players, 
we designed \emph{\GFi{}} to turn off the green spotlights of open players beyond a pre-defined filter radius.
\GFi{} incorporates three considerations:

\begin{itemize}[topsep=0pt]
    \item \textbf{Trigger}:
    Generally speaking, the system should always avoid overwhelming the user.
    Thus, \GFi{} is consistently triggered
    and updated when the user moves his/her gaze.
    It centers at the user's gaze point with a filter radius of 650px (empirically selected). 
    
    \item \textbf{Visual feedback}:
    To indicate the user that the interaction is being triggered, 
    we designed a radial blurring effect 
    that darkens the audience outside the filter radius
    and updates dynamically.
    We smooth the movement of the radial blurring effect
    to prevent it from abruptly changing location due to the user's saccade.
    Note that the blurring effect will not be applied to players and the court to ensure their readability.
    This visual feedback can notify the user about the existence of the interaction
    while also creating a theater mode that helps the user to focus on and engage with the game.
    
    \item \textbf{Outcome}: 
    The green spotlights, 
    which are used to highlight offensive players with open spaces, 
    outside the filter radius will be turned off (\autoref{fig:gaze_int}b).
    An ease-in-out effect is applied to the change.

\end{itemize}
\section{User Study}
\label{sec:evaluation}

To assess the usefulness, usability, and engagement 
of using our gaze-moderated embedded visualizations in watching basketball videos,
we conducted a comparative study between three modes --
watch with raw footage (\RAW{}), 
with solely embedded visualizations (\AUG{}),
and with gaze-moderated embedded visualizations (\FULL{}).

\begin{figure*}[hb]
    \centering
    \includegraphics[width=0.95\textwidth]{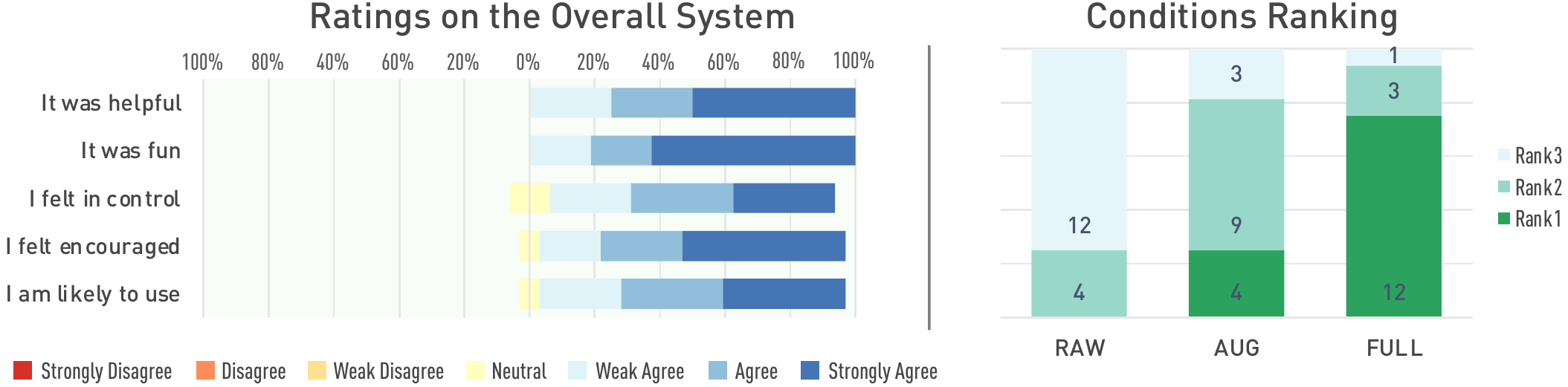}
    \caption{Left: Ratings on the overall system. Right: Rankings of different modes, 
        i.e., \RAW{} -- watch with raw footage,
        \AUG{} -- watch with embedded visualizations solely,
        \FULL{} -- watch with gaze interaction and embedded visualizations.
        }
    \label{fig:overall}
\end{figure*}

\subsection{Study Setup}
\label{sec:study_setup}

\subsubsection{\textbf{Participants and Apparatus}}
\noindent
We recruited 16 casual fans (F1-F16; M = 10, F = 6; Age: 18 - 35) through university mailing lists and forums 
after screening for their fandom levels and basketball knowledge.
All participants
watched \quot{less than 10 games per year} 
and only \quot{know the basic rules} of basketball.
All participants had not participated in our formative study.
The study was conducted in the lab with a 24-inch monitor.
We followed the same ergonomic settings in the formative study 
but changed to use a Tobii eye tracker 5 (\cmo{133Hz})~\cite{tobii} 
for more accurate gaze interactions.
The study took about one hour and each participant was compensated with a \$20 gift card.

\subsubsection{\textbf{Design and Procedure}}
\noindent
The study consisted of two tasks, namely,
\textbf{Task1 - \RAW{} vs. \AUG{}}, 
and \textbf{Task2 - \AUG{} vs. \FULL{}}.
Each task compared two modes.
For each task,
we used a game video from the formative study
and evenly split it into two video clips (each lasting around 4.5 minutes)
for each mode.
The videos and the order of modes in each task
were counterbalanced across participants.
Each session included the following phases:

\paragraph{Phase 1. Introduction (10mins).}
The study started with an introduction of the research motivation, 
the purpose of the study, 
and the protocol.
After the participant signed a consent form, 
we conducted a warm-up interview about basketball game-watching experiences.

\paragraph{Phase 2. Comparative Tasks (40mins).}
We asked participants to finish two comparative tasks (each lasted 20 mins):
In \textbf{Task1 - \RAW{} vs. \AUG{}}, 
the participants watched two video clips in \RAW{} and \AUG{} modes, respectively.
    Before \AUG{} mode, we conducted a training session to
    walk the participant through the four embedded visualizations in a separate video (about 20 seconds).
    We only proceeded to the task when participants were clear and confident enough to use the embedded visualizations.
In \textbf{Task2 - \AUG{} vs. \FULL{}}, 
the participants watched another two video clips in \AUG{} and \FULL{} mode, respectively. 
Again, a training session was conducted before starting \FULL{} mode to ensure the participant were confident to use the gaze interactions.
Participants were encouraged to think aloud about their game observations when watching the videos.
At the end of each video clip,
we performed a post-video interview to
collect the participants' feedback on the mode they had just experienced.
At the end of each task, 
participants filled out a post-task questionnaire to rate their experiences.

\paragraph{Phase 3. Post-study Questionnaire (10mins).}
We asked participants to complete a post-study questionnaire of 
their subjective ratings on the overall system,
rank the three modes,
and provide feedback on the entire system.

\subsubsection{\textbf{Measures}}
\noindent
We collected quantitative measurements of user subjective ratings 
in the post-task 
and post-study questionnaires.
At the end of each task, 
participants were asked to rate the usefulness, engagement, and usability of the features 
they had just experienced,
including the four visualizations 
(i.e., Player Highlight, \OffR{}, \DefS{}, and \OOOL{}) in \textbf{Task1}
and the two gaze interactions 
(i.e., \GFo{} and \GFi{}) in \textbf{Task2},
on a 7-point Likert scale.
In the post-study questionnaires,
we asked participants to rate the overall system on five questions about system usefulness and engagement~\cite{obrienDevelopmentEvaluationSurvey2010} 
and ranked the three modes. 

\subsection{Study Results}
We first report 
the ratings of the overall system 
and the rankings of the three modes,
and then discuss the feedback on the usefulness, engagement, and usability of
individual visualizations and interactions.

\subsubsection{\textbf{The overall user experience of \system{} was predominantly positive, with \FULL{} being most preferred}}

\noindent
Figure~\ref{fig:overall} left shows that
the majority rated \system{} as \quot{helpful} and \quot{fun}, 
felt \quot{in control} and \quot{encouraged} when using the system,
and were \quot{likely to use} it for watching basketball games.
Figure~\ref{fig:overall} right presents the rankings of the three modes,
showing that 
\FULL{} was the most preferred mode by 12 participants, 
followed by \AUG{} by 4 and \RAW{} by none. 
The four participants who didn't rank \FULL{} as the best
were mainly concerned about  
the blurring effect of the audience in \GFi{}, 
stating that 
a game video without audiences seemed abnormal.
However, they agreed that the filtering of open players (highlighted in green) was useful.
Thus, the system should allow users to turn off the blurring effect. 

\begin{figure*}[h]
    \centering
    \includegraphics[width=0.99\textwidth]{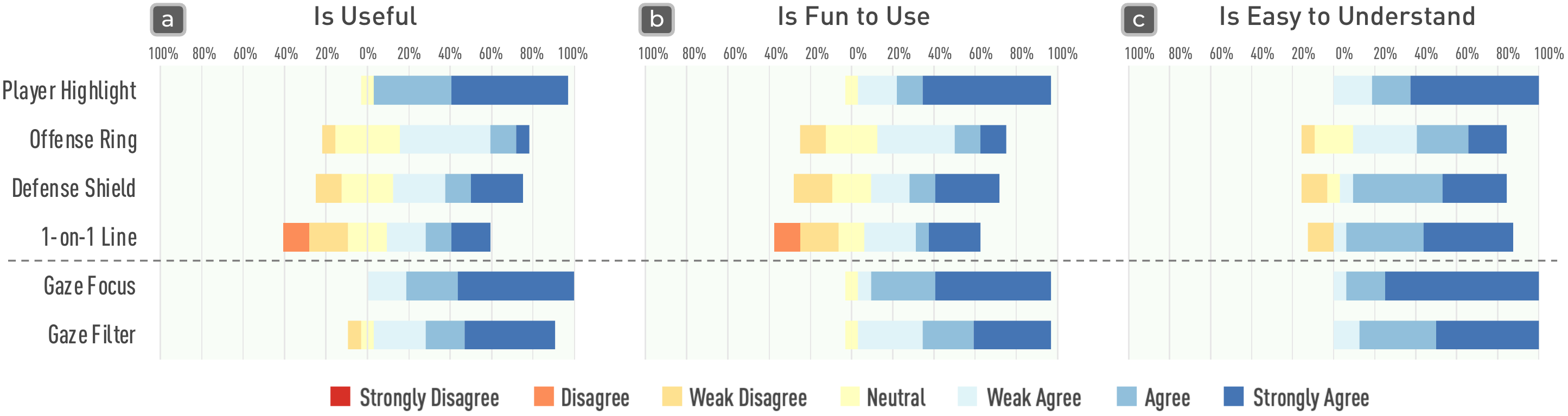}
    \caption{Ratings on the usefulness, engagement, and usability of the embedded visualizations and gaze interactions in \system{}.}
    \label{fig:rate_vis}
\end{figure*}

\subsubsection{\textbf{Embedded visualizations are more useful if they are more attractive and informative}}
\noindent
Participants rated positively on the usefulness of each visualization (\autoref{fig:rate_vis}a).
Among the four embedded visualizations,
Player Highlight was considered to be the most useful
in \quot{predicting the ball receiver} at the next pass,
\quot{highlighting the open players},
and thus \quot{making the game much clearer}, 
as mentioned by the participants.

Participants also rated \OffR{} and \DefS{} as useful to understand players' in-game decisions.
For example, F8 thought that \OffR{}
could \quot{help me expect when he's gonna shoot. Otherwise, it's just like this person can shoot anytime.};
F3 commented that \DefS{} clearly showed \quot{who's able to defend} 
and explained why \quot{James can easily score on Iguodala}.
However, 
some participants found them less useful 
since the visualizations were placed on the ground,
different from the ball at hand-height,
and thus often hard to notice in a tense game.
Nevertheless, most of the participants agreed that 
\quot{it was nice to have} \OffR{} and \DefS{} in the system.

\OOOL{} was controversial in terms of usefulness. 
Some participants 
considered it helpful \quot{to follow the game and to see who was involved [in defending]} (F6)
while others felt it suffers from the same limitations 
as \OffR{} and \DefS{} (i.e., cannot be noticed)
with less useful information 
that was already \quot{clear from the video.} (F1)

\subsubsection{\textbf{Gaze interactions are useful 
as they satisfy audiences' personal information and cognitive needs
}}
\noindent
Both \GFo{} and \GFi{}
received very positive ratings regarding usefulness (\autoref{fig:rate_vis}a).
All participants enjoyed using \GFo{}
as it could help them immediately \quot{know the name of the player [who I am looking at]} with simple and effective interaction.
\GFi{} also 
improved the game understanding of participants
as it could \quot{make the scene tidier} (F2)
and \quot{focus the information on what I'm looking [at].} (F4)

\subsubsection{\textbf{Embedded visualizations are engaging since they provide a deeper epistemic pleasure}}
\noindent
The participants felt that 
the embedded visualizations were engaging (\autoref{fig:rate_vis}b)
since they could help them \quot{understand and follow the game}.
This is also reflected 
in the positive relationship between the usefulness and engagement of the embedded visualizations -- 
the more useful, the more engaging.
For example,
F1 
commented that \quot{it is nice to know who the star players are because otherwise, they all look like regular players to me.}
F5 remarked that he \quot{always tried to predict the receiver of a pass but always failed.}
With the visualizations, he felt much more confident in predicting the receiver and 
felt quite a sense of accomplishment whenever he was correct.

\subsubsection{\textbf{Gaze interactions are engaging since they promote proactive game viewing experiences}}
\noindent
Gaze interactions are also engaging (\autoref{fig:rate_vis}b)
since they make the passive watching experience
interactive and proactive.
F2 said when the video scene responded to her gaze, 
she felt that she was \quot{a part of the game.}
F4 particularly enjoyed using the gaze interactions, 
which made him \quot{almost feel like I'm there.}
Moreover, participants provided that using the gaze interactions was \quot{natural} 
and would not add further cognitive cost to the game watching 
since \quot{you can actively control it ... you don't have to [use it]} (F5).
Overall, 
the participants' feedback
provides a strong hint 
that the engagement of watching sports videos can be improved
if video content responds to the audience's gaze.

\subsubsection{\textbf{The gaze-moderated embedded visualizations are easy to understand and use}}
\noindent
All participants confirmed the usability of the embedded visualizations and gaze interactions in \system{} (\autoref{fig:rate_vis}c).
They thought both the visualizations and gaze interactions were \quot{easy to understand} and \quot{use}.
We noticed that while some participants thought that \OOOL{} was not that useful,
they still agreed that it was easy to understand.

\subsection{Implications for Designing Gaze-moderated Embedded Visualizations}
We now discuss the design implications we learned from the feedback and observations in the user study.

\subsubsection{\textbf{Visual attention matters when designing embedded visualizations}}
\noindent
Instead of being overwhelmed by the embedded visualizations,
some participants sometimes even could not notice several of them. 
This may be due to a phenomenon known as 
Inattentional Blindness~\cite{DBLP:journals/tvcg/HealeyE12},
in which viewers can fail to perceive visually salient objects or activities.
This finding implies that 
designers should consider \textbf{how to properly direct the user's attention when designing embedded visualizations} to effectively convey information and avoid overwhelming the user.
For instance,
F5 suggested that 
we should highlight \OffR{}, instead of players,
to help audiences efficiently identify the player with the highest EPV;
Highlighting visualizations that are linked to immediate actions
can increase the information salience, 
e.g., highlighting the \OffR{} when a player is about to shoot.

\subsubsection{\textbf{Synchronization of seeing and hearing matters when designing embedded visualizations}}
\noindent
F4 mentioned that he paid less attention to the commentaries in \AUG{} mode
but listened to them more carefully in \RAW{} mode,
which implied that the embedded visualizations could ``overlay'' the commentaries.
In \FULL{} mode, 
several participants reported using gaze interactions to search for players mentioned in the commentaries. 
These observations indicate that
there is an interaction between 
the participants' perception of the embedded visualizations and the commentaries.
Such an interaction between vision and hearing has long been identified (e.g., McGurk effect~\cite{tiippana2014mcgurk}).
This highlights the importance of \textbf{synchronizing the embedded visualizations and the commentaries to create a consistent watching experience}.
Some participants suggested leveraging the commentaries to create embedded visualizations, as explored in a recent paper~\cite{chen2022sporthesia}.


\subsubsection{\textbf{Gaze interactions shift the game from explanatory to exploratory}}
\label{sec:explan_2_explor}

\noindent
In \FULL{} mode,
the participants spent more time using their gaze to highlight players,
while in \AUG{} mode,
participants tended to follow the players highlighted by the system.
This difference suggests 
that \AUG{} mode is more explanatory
while the addition of gaze interactions can shift it towards exploratory.
This is not surprising,
as the gaze interactions allows the audiences to actively explore the game more.
When designing gaze interactions for game viewing systems or, broadly speaking, 
any situated visualization systems that involve visual guidance,
designers must \textbf{consider the ultimate goal of the systems
and strike a balance between explanatory and exploratory}.

\subsubsection{\textbf{Gaze interactions enable active learning in game watching}}
\noindent
The gaze interactions can 
also help audiences learn basketball knowledge progressively.
For example, 
\GFi{} only highlights \cmo{open} players with green spotlights when the players are near the user's gaze.
F3 used this feature to verify his hypotheses of team tactics
by moving his gaze to some areas and seeing
if the system \quot{showed green [highlighting]} there.
F15 elaborated that \GFo{}
helped him better recognize players
by showing the name of a player to confirm 
that he was looking at the right person.
Such a hypothesis-testing process 
made the participants feel more confident in 
interpreting the game. 
We see this as an interesting opportunity 
to \textbf{leverage gaze-moderated embedded visualization 
to develop long-term impact for the users}
beyond improving their watching experience within individual games.

\subsubsection{\textbf{Suggestions}}
\noindent
While the participants generally spoke highly of \system{},
they did mention a few limitations related to the system implementations.
For example, F8 disliked the visual artifacts introduced by the imperfect segmentation model;
some participants said that \GFo{} was not accurate when the players crowded together.
These limitations can potentially be resolved with more advanced models or eye trackers.
The participants also suggested several improvements for \system{},
such as 
providing more customization options (e.g., for the visualizations and gaze interactions) through a \quot{Preference} panel 
and allowing the visualizations to adapt to the pace of the game (e.g., e.g., showing more details in slow-paced and less in fast-paced situations).
Another point worth mentioning is that 
a few participants wished the system could generate play-by-play replays with embedded visualizations
to explain the game in detail.
This could be an interesting direction for future research.

\subsection{\cmo{Feedback from Broader Users}}
\cmo{
While \system{} is designed for casual fans, 
we also conducted a follow-up study with die-hard fans 
to explore its potential use beyond our original target users. 
We recruited 8 die-hard fans (D1-D8; M = 8; Age: 18 - 35),
who knew \quot{basketball tactics and pros and cons of specific players} 
and watched \quot{at least 1 game per week}. 
No female die-hard fan responded to us.
We followed the same process as in \autoref{sec:study_setup}
to help the die-hard fans experience our system 
with a focus on collecting feedback on the real-world use of \system{}.
We discuss their major opinions that differed from those of the casual fans, as well as how \system{} can be further improved.
}

\subsubsection{\cmo{\textbf{\system{} can improve game understanding and engagement for die-hard fans}}}
\noindent
\cmo{
All die-hard fans confirmed the usefulness of \system{} in watching basketball videos.
Unlike the casual fans (F1-F16), the die-hard fans could gain a deeper understanding of the game with the embedded visualizations.
For example, they could further recognize the offensive tactics of the team from the highlighted open players.
For some die-hard fans, the usefulness of \system{} extends beyond understanding the games.
D4 - D7, for example, were basketball players themselves and believed that \system{} could help them improve their in-game decisions and tactics.
On the other hand, the die-hard fans did request extra in-depth data that were currently not supported by \system{}, 
such as the trajectories of the players' off-ball movements.
Besides game understanding, the die-hard fans also agreed that \system{} could enhance their engagement in game watching,
especially the gaze interactions, which gave them a feeling of \quot{participating in the game.} (D1)
}

\subsubsection{\cmo{\textbf{Customization is indispensable for the die-hard fans}}}
\noindent
\cmo{
The feedback from the die-hard fans indicates that there is no one-size-fits-all design that will satisfy everyone. 
Compared to the casual fans, the die-hard fans had more diverse opinions on the features.
For instance, 
D5 preferred highlighting fewer open players, while D6 preferred highlighting more;
D1 thought that showing the names could help him learn about unfamiliar players,
while some participants only cared about the ``star'' players;
D7 and D2 wanted more gaze interactions, but D4 found them distracting.
While their preferences vary,
they all agreed that there are valid reasons for the different design choices
and that the best solution is to give users the option to customize the system, which is aligned with findings by Lin et al.~\cite{lin2022the}.
One interesting question is how to help users efficiently express their preferences for customization, as the range of possible configurations can be very large.
}

\subsubsection{\cmo{\textbf{Embedded visualizations do not need to be displayed throughout the entire game}}}
\noindent
\cmo{
Seven out of 8 die-hard fans thought that they did not need the embedded visualizations to be displayed throughout the entire game.
They explained that \system{} is most useful when players are executing the coach's strategy, such as in the first two quarters.
However, 
when the game is decided by the ``star'' players' in-game states and improvisations (during crunch time), 
the embedded visualizations may be less useful.
In addition, some participants (e.g., D4, D5) felt that watching a full game with gaze interactions can be exhausting,
as they would \quot{keep using the interactions}.
This echoes our finding in \autoref{sec:explan_2_explor} that gaze interactions promote proactive analysis.
The participants suggested that the system should allow users to decide when to display the embedded visualizations.
}


\subsubsection{\cmo{\textbf{Gaze interactions should provide more adaptive data for the die-hard fans}}}
\noindent
\cmo{
When using the gaze interactions,
the die-hard fans wanted more adaptive data for different teams, players, and game events.
For example, 
the retrieved data for the Golden State Warriors could focus on teamwork, 
while the data for the Cleveland Cavaliers could emphasize the performance of their ``star'' players.
When gazing two ``star'' players facing off against each other, 
such as LeBron James vs. Steph Curry, 
the system could display their historical one-on-one records.
Besides, the data could adapt to specific game events, such as dunking, or the intensity of the game.
These suggestions, which would require a more intelligent and sophisticated system, are left for future research.
}
\section{Discussions}

\cmo{
In this section, we will discuss potential future research directions and limitations of our current study.}

\paragraph{\textbf{Reproducible Environments For Embedded Visualizations Research.}}
Compared to traditional web-based visualizations,
embedded or situated visualizations are particularly challenging to research
since the physical context where they are registered in
is inherently difficult to reproduce, distribute, and benchmark.
It can be more difficult, or even impossible, to reproduce
the physical context if it is dynamic (e.g., sports scenarios).
This perhaps is the major reason why most existing research~\cite{lin2022the, yao2022visualization}
uses reproducible simulated environments (e.g., virtual reality) 
to study embedded visualizations.
In this work, 
we use videos to explore 
the design of interactive embedded visualizations 
in dynamic, complex scenarios.
To advance research in embedded visualizations,
we will open source
our video-based environments (i.e., code and data)
so that others can
reproduce our system and develop their own.

\vspace{-0.5mm}
\paragraph{\textbf{Gaze Interactions for Embedded Visualizations.}}
In recent years, eye-tracking technology has become increasingly affordable. 
Compared to other input modalities such as keyboard, mouse, and voice, gaze input can enable fast, intuitive, and implicit interactions.
In fact,
gaze interactions are widely supported in head-mounted displays~\cite{vive2}
for augmented or virtual reality (AR/VR),
which are the main scenarios for using embedded visualizations.
Our research shows that even only using simple gaze data (i.e., the 2D position of the gaze point on the screen)
can significantly increase the usefulness and engagement of embedded visualizations.
However, 
gaze interactions have their own limitations,
such as \cmo{that they} cannot be used in multi-viewers scenarios (e.g., TV in living rooms).
Besides,
we have not yet explored 
using advanced gaze events \cmo{(e.g., fixation, saccade, pursuit)}
or combining gaze with other input modalities \cmo{(e.g., speech)}
\cmo{to achieve more adaptive or customized embedded visualizations},
which we consider as promising future directions.

\vspace{-0.5mm}
\paragraph{\textbf{Towards Augmenting Live Game Viewing.}}
In \autoref{sec:cv_limitation}, 
we discussed the technical challenges and potential solutions for extending the CV pipeline to livestreams.
\cmo{Additionally,
to support the embedded visualizations in livestreams,
the system also requires real-time tracking data of the players.
If this data is unavailable,
a potential solution is to use camera calibration techniques~\cite{Sha_2020_CVPR, DBLP:journals/tmm/HuCWC11} to estimate the camera parameters,
which can be used to estimate the players' positions and calculate the offensive and defensive metrics.
}
On the other hand,
real-world scenarios also provide additional resources
for improving the system,
including 
videos with a higher resolution and framerate,
buffer time in streaming,
camera parameters, 
and steering from human experts.
Thus, 
we believe our system can be extended 
to live game videos by the broadcasters
or researchers once these additional resources are available.

\vspace{-0.5mm}
\paragraph{\textbf{Augmenting Real-world Games Beyond Videos.}}
With the development of sensing techniques and AR devices 
such as head-mounted displays,
it becomes increasingly possible to augment 
real-world games with digital information.
While our research provides a step-stone towards augmenting real-world environments,
several issues must be taken into account when adopting it to AR,
including the limited field of view, 
the effect of stereoscopy vision and depth perception,
and the ability to freely change viewing perspective.
We hope that the lessons learned from the present research can 
inspire and provide a solid foundation for future research on augmenting in-person game watching scenarios, 
ultimately generalizing embedded visualizations to general real-world environments.

\vspace{-0.5mm}
\paragraph{\textbf{Study Limitations.}}
Our user study 
only evaluated the system on \cmo{G1 and G2 rather than videos of entire basketball games.}
The study results, including the ratings and gaze distributions,
only provide qualitative evidences.
The designs of the embedded visualizations and gaze interactions in \system{}
are derived based on the 16 participants in our formative study.
Further explorations are thus suggested for different scenarios and user groups.
\section{Conclusion}
This work explores using gaze-moderated embedded visualizations 
to facilitate game understanding and engagement of casual fans.
We compared the game-watching behaviors of casual and die-hard fans in a formative study
to identify the particular pain points of casual fans in watching basketball videos.
Based on the findings,
we developed a CV pipeline to support \system{},
a basketball video-watching system
equipped with gaze-moderated embedded visualizations.
With \system{}, casual fans 
can effectively identify key players,
understand their in-game decisions,
and personalize the game-viewing experiences through natural gaze interactions.
\cmo{
We evaluated the CV pipeline with computational experiments.
A user study with 16 casual fans confirmed
the usefulness, engagement, and usability of \system{}.
We further collected feedback on \system{} from 8 die-hard fans.
The feedback of these 24 participants provides useful suggestions to improve \system{} 
and insightful implications for future research in interactive embedded visualizations for sports game viewing.
}

\newpage

\begin{acks}
The authors wish to thank Salma Abdel Magid for her beautiful voice and help on the video narration.
This research is
supported in part by the 
NSF award III-2107328, NSF award IIS-1901030,
NIH award R01HD104969,
and the Harvard Physical Sciences
and Engineering Accelerator Award.
\end{acks}

\bibliographystyle{ACM-Reference-Format}
\bibliography{bibtex}

%
\newpage
\appendix
\label{app:A}

\section{Player Dataset}


\begin{table}[h]
    \centering
    \caption{Statistics for our player dataset.}
    \label{tab:my_label}
    \begin{tabular}{l|c|c}
        \toprule
             & \textbf{Num of Frames} & \textbf{Num of Labels} \\ \midrule
        \textbf{G1 - train} & 9552 & 8837 \\ \hline
        \textbf{G1 - test} & 4029 & 3596 \\ \hline
        \textbf{G2 - train} & 8312 & 6530 \\ \hline
        \textbf{G2 - test} & 3485 & 3485 \\\bottomrule
    \end{tabular}
\end{table}

\begin{table}[htb]
        \centering
        \captionsetup{width=.8\textwidth}
        \vspace{0pt}
        \caption{Number of labeled instances for each player class in G1.}
       \label{tab:my_label2}
        \begin{tabular}{l|l}
            \toprule
            \textbf{Class Label} & \textbf{\# of instances} \\ \hline
            G30 & 12574 \\
            G9 & 12953 \\
            G23 & 12404 \\
            C0 & 12418 \\
            C8 & 12461 \\
            C23 & 12354 \\
            G11 & 11760 \\
            G34 & 11834 \\
            C4 & 10629 \\
            C2 & 6663 \\
            C5 & 5268 \\
            C13 & 1508 \\
            G31 & 643 \\
            \bottomrule
        \end{tabular}
\end{table}

\begin{table}[htb]
\captionsetup{width=.8\textwidth}
    \vspace{0pt}%
        \centering
        \caption{Number of labeled instances for each player class in G2.}
       \label{tab:my_label3}
        \begin{tabular}{l|l}
            \toprule
            \textbf{Class Label} & \textbf{\# of instances} \\ \hline
            O0 & 10375 \\
            L3 & 11115 \\
            O9 & 9761 \\
            O3 & 9407 \\
            O35 & 9194 \\
            L6 & 9774 \\
            L32 & 9401 \\
            O12 & 9235 \\
            L1 & 5576 \\
            L11 & 5632 \\
            L12 & 3864 \\
            L4 & 3632 \\
            O21 & 653 \\
            L33 & 642 \\
            O2 & 506 \\
            \bottomrule
        \end{tabular}
\end{table}

\noindent 
We created a dataset for the two videos (G1 and G2) used in our formative study.
For both videos, 
we first removed all transition scenes (e.g., replays)
since transition scenes typically show close-up views of the players
and can be noise for the detector.
We gathered 13581 and 11797 frames for G1 and G2, respectively.
For each frame,
we identified the players with at least half of the body in the scene
and labeled their bounding boxes with the players' unique IDs. 
Occluded players were also labeled if at least half of their bodies were in the scene.
In total, there are 13 and 15 unique IDs (i.e., classes) in G1 and G2, respectively.
Table.~\ref{tab:my_label} shows an overview statistics of our dataset.
Table.~\ref{tab:my_label2} and Table.~\ref{tab:my_label3} provide a detailed breakdown of the number of labeled instances the datasets have for each class. 
The first letter of the class names (except Negative) indicates the player's team 
(where G = Golden State, C = Cleveland, O = Oklahoma City, and L = Los Angeles), 
and the number that follows is the number of the player's labels.

\end{document}